\documentclass[proceedings]{JHEP37}
\usepackage{eurosym}
\usepackage{amssymb}
\usepackage{amsfonts}
\usepackage{amsmath,epsfig}
\usepackage{graphicx}

\newbox\mybox

\newcommand\fverb{\setbox\mybox=\hbox\bgroup\verb}
\newcommand\fverbdo{\egroup\medskip\noindent\fbox{\unhbox\mybox}\ }
\newcommand\fverbit{\egroup\item[\fbox{\unhbox\mybox}]}
\conference{Complex solitons with real energies}
\abstract{Using Hirota's direct method and B\"{a}cklund transformations we construct explicit
complex one and two-solutions to the complex Korteweg-de Vries equation, the complex modified Korteweg-de Vries
equation and the complex sine-Gordon equation. The one-soliton solutions of trigonometric and elliptic type turn out to be $\mathcal{PT}$-symmetric when a constant of integration is chosen to be purely imaginary with one special choice corresponding to solutions recently found by Khare and Saxena. We show that alternatively complex $\mathcal{PT}$-symmetric solutions to the Korteweg-de Vries equation may also be constructed alternatively from real 
solutions to the modified Korteweg-de Vries by means of Miura transformations. The multi-soliton solutions
obtained from Hirota's method break the $\mathcal{PT}$-symmetric, whereas those obtained from B\"{a}cklund transformations are $\mathcal{PT}$-invariant under certain conditions. Despite the fact that some of the Hamiltonian densities are non-Hermitian, the total energy is found to be positive in all cases, that is irrespective of whether they are $\mathcal{PT}$-symmetric or not. The reason is that the symmetry can be restored by suitable shifts in space-time and the fact that any of our N-soliton solutions may be decomposed into N separate $\mathcal{PT}$-symmetrizable one-soliton solutions.}

\title{Complex solitons with real energies}
\author{Julia Cen and Andreas Fring \\
Department of Mathematics, City University London,\\
Northampton Square, London EC1V 0HB, UK \\
E-mail: julia.cen.1@city.ac.uk, a.fring@city.ac.uk}

\input{tcilatex}
\begin{document}

\section{Introduction}

$\mathcal{PT}$-symmetrically deformed nonlinear wave equations have been
found to possess various interesting properties \cite%
{BBCF,AFKdV,Josh,BBAF,PEGAAFint,comp1,CompPaulo,CFB,YanZ,ACAF2,assis2015pt}.
In general, the $\mathcal{PT}$-symmetric deformations destroy the
integrability of models with that property, although some rare cases pass
the Painlev\'{e} test \cite{PEGAAFint} indicating that they remain
integrable. Furthermore, it was shown \cite{BBAF} that it is possible to
construct specific $\mathcal{PT}$-symmetric deformations that preserve the
supersymmetry of some models. Some $\mathcal{PT}$-symmetrically deformed
nonlinear wave equations possess very intricate shock wave structures \cite%
{ACAF2}.

Most notably when the $\mathcal{PT}$-symmetrically deformed models are of
Hamiltonian type, with densities ${\mathcal{H}}[u(x,t),u_{x}(x,t),\ldots ]$
depending on some field $u(x,t)$ and its derivative, the $\mathcal{PT}$%
-symmetry will ensure that the energy on symmetric intervals $[-a,a]$%
\begin{equation}
E=\int\nolimits_{-a}^{a}{\mathcal{H}}[u(x,t)]dx=\oint\nolimits_{\Gamma }{%
\mathcal{H}}[u(x,t)]\frac{du}{u_{x}},  \label{E}
\end{equation}%
remains real despite the fact that the Hamiltonian density is complex \cite%
{AFKdV}. The reasoning to establish this is similar to the one applied to
quantum mechanical models, although the quantity $E$ as defined in (\ref{E})
will not play the role of the energy in the quantum theory for reduced
models as explained in more detail in \cite{CFB}.

The simplest way to obtain complex solutions is to keep the form of the
original equation intact and just take the field $u(x,t)$ to be complex by
demanding that the complexified equations remain invariant under the
antilinear transformation $\mathcal{PT}$: $x\rightarrow -x$,\ $t\rightarrow
-t$, $i\rightarrow -i$ and $u\rightarrow u$ or $u\rightarrow -u$. For such a
setting Khare and Saxena \cite{khare2015novel} have recently found some
interesting apparently novel $\mathcal{PT}$-symmetric solutions to various
types of nonlinear equations that appear to have been overlooked this far.
Their approach is to start off from some well-known real solutions to these
equations and then by adding a term build around that solution a suitable
complex Ansatz including various constants. In many cases they succeeded to
determine those constants in such a way that their expressions constitute
solutions to the different types of complex nonlinear wave equations
considered.

One of the purposes of this note is to demonstrate that these solutions may
be derived in a more constructive, systematic and generic way. We focus here
on nonlinear wave equations for which we use Hirota's direct method \cite%
{Hirota} to derive complex solutions including those of \cite{khare2015novel}
as special cases. Some of the one-soliton solutions produced in this manner
turn out to be $\mathcal{PT}$-symmetric, whereas the multi-soliton solutions
obtained from this method break the $\mathcal{PT}$-symmetry in general.
Subsequently we employ B\"{a}cklund transformations to construct new $%
\mathcal{PT}$-symmetric multi-soliton solutions from some previously
constructed complex solutions. For a specific case we evaluate the
time-delay in the real and imaginary parts of these solutions.

Computing the energies $E$ corresponding to our solutions we find that all
of them are real irrespective of whether they are $\mathcal{PT}$-symmetric
or not. While this is to be expected for the $\mathcal{PT}$-symmetric
solutions, this is less obvious for the $\mathcal{PT}$-broken solutions. We
will present the argument and mechanism responsible for this behaviour. Our
analysis is carried out for the complex Korteweg-de Vries (KdV) equation in
section 2.1, complex modified Korteweg-de Vries (mKdV) equation in section
2.2, both considered also in \cite{khare2015novel}, and in addition for the
complex sine-Gordon equation in section 2.3. Our conclusions are stated in
section 3.

\section{The construction of complex multi-soliton solutions}

At first we employ here Hirota's direct method \cite{Hirota}. The general
principle of this approach is to convert the nonlinear equation of interest
into Hirota's equation of bilinear form by means of a dependent variable
transformation%
\begin{equation}
P(D_{1},D_{2},\ldots ,D_{n})\tau \cdot \sigma =0,
\end{equation}%
with $P(D_{1},D_{2},\ldots ,D_{n})$ being a polynomial in the Hirota
derivatives $(D_{1},D_{2},\ldots ,D_{n})$ acting on the product of the two
functions $\tau $ and $\sigma $ both depending on $(x_{1},x_{2},\ldots
,x_{n})$. The general expressions for the Hirota derivatives in terms of
ordinary derivatives may be obtained from the generating function%
\begin{equation}
\tau (x_{1}+y_{1},\ldots ,x_{n}+y_{n})\sigma (x_{1}-y_{1},\ldots
,x_{n}-y_{n})=e^{y_{1}D_{1}+y_{2}D_{2}+\ldots +y_{n}D_{n}}\tau \cdot \sigma ,
\end{equation}%
by reading off powers in $y_{i}$. In particular, we shall require below the
expressions%
\begin{eqnarray}
D_{t}\tau \cdot \sigma &=&\tau _{x}\sigma -\sigma _{x}\tau ,  \label{H1} \\
D_{x}^{2}\tau \cdot \sigma &=&\tau _{xx}\sigma -2\tau _{x}\sigma _{x}+\tau
\sigma _{xx},  \label{H2} \\
D_{x}^{3}\tau \cdot \sigma &=&\tau _{xxx}\sigma -3\tau _{xx}\sigma
_{x}+3\tau _{x}\sigma _{xx}-\tau \sigma _{xxx},  \label{H3} \\
D_{x}^{4}\tau \cdot \sigma &=&\tau _{xxxx}\sigma -4\tau _{xxx}\sigma
_{x}+6\tau _{xx}\sigma _{xx}-4\tau _{x}\sigma _{xxx}+\tau \sigma _{xxxx},
\label{H4} \\
D_{x}D_{t}\tau \cdot \sigma &=&\tau _{xt}\sigma +\tau _{xt}\sigma -\tau
_{x}\sigma _{t}-\tau _{t}\sigma _{x}.  \label{H5}
\end{eqnarray}%
The solution procedure is then to expand the functions $\tau $ and $\sigma $
in powers of $\lambda $ as $\tau =\sum\nolimits_{k=0}^{\infty }\lambda
^{k}\tau ^{k}$, $\sigma =\sum\nolimits_{k=0}^{\infty }\lambda ^{k}\sigma
^{k} $ and subsequently solve the \emph{bilinear} Hirota equation order by
order in $\lambda $. It turns out that one can systematically set $\tau
^{k}=\sigma ^{k}=0$ for some $j\leq k$. The constant $\lambda $ may then be
absorbed into the $\tau ^{k}$ and $\sigma ^{k}$ so that the terminated
series constitute an \emph{exact} solution to the Hirota equation and
therefore, after re-transformation, to the original nonlinear equation.

\subsection{The complex Korteweg-de Vries equation}

The KdV equation for the complex field $u(x,t)$ may be considered as a set
of coupled equations for the real fields $p(x,t)$ and $q(x,t)$ 
\begin{equation}
u_{t}+6uu_{x}+u_{xxx}=0\quad \Leftrightarrow \quad \left\{ 
\begin{array}{r}
p_{t}+6pp_{x}+p_{xxx}-6qq_{x}=0 \\ 
q_{t}+6\left( pq\right) _{x}+q_{xxx}=0%
\end{array}%
\right. ,  \label{KdV}
\end{equation}%
when taking $u=p+iq$. The coupled equations reduce to the Hirota-Satsuma 
\cite{hirota1981soliton} and Ito system \cite{ito1982symmetries} when
setting $\left( pq\right) _{x}\rightarrow pq_{x}$ and $q_{xxx}\rightarrow 0$
in the second equation, respectively. Evidently these equations remain
invariant for $\mathcal{PT}$: $x\rightarrow -x$,\ $t\rightarrow -t$, $%
i\rightarrow -i$, $u\rightarrow u$, $p\rightarrow p$, $q\rightarrow -q$. We
stress here that, although there are many $\mathcal{PT}$-symmetric solutions
to (\ref{KdV}), not all solutions to (\ref{KdV}) need to be $\mathcal{PT}$%
-symmetric since the symmetry could map one solution, say $u_{1}(x,t)$, into
a new one $u_{1}(-x,-t)=u_{2}(x,t)\neq u_{1}(x,t)$. Unlike as in the linear
quantum mechanical scenario the sum of these two solution would of course
not constitute a new $\mathcal{PT}$-symmetric solution, as the KdV equation
is nonlinear. In fact, it would not be a solution at all, unless $%
(u_{1}u_{2})_{x}=0$.

\subsubsection{Complex solutions from the Hirota method}

Since the original work of Hirota \cite{Hirota} it is well known that the
KdV equation (\ref{KdV}) can be converted into Hirota's bilinear form 
\begin{equation}
\left( D_{x}^{4}+D_{x}D_{t}\right) \tau \cdot \tau =0,  \label{Hirota}
\end{equation}%
by means of the variable transformation $u=2(\ln \tau )_{xx}$ together with (%
\ref{H4}) and (\ref{H5}). Equation (\ref{Hirota}) is solved easily with the
above mentioned expansion for $\tau $. At order $\lambda ^{0}$ the equation
is trivially satisfied and at order $\lambda ^{1}$ we have to solve 
\begin{equation}
\left( D_{x}^{4}+D_{x}D_{t}\right) (1\cdot \tau ^{1}+\tau ^{1}\cdot
1)=2(\tau ^{1})_{xt}+2(\tau ^{1})_{xxxx}=0.  \label{zwei}
\end{equation}%
Thus the original problem to solve a nonlinear equation has been reduced to
the much simpler task of just solving a linear equation. We may now take 
\begin{equation}
\tau ^{1}=e^{\eta _{1}}\qquad \text{with }\eta _{i}=k_{i}x+\omega _{i}t+\mu
_{i}\text{, }k_{i},\omega _{i}\in \mathbb{R}\text{, }\mu _{i}\in \mathbb{C}%
\text{,}  \label{etac}
\end{equation}%
with nonlinear dispersion relation $k_{1}^{3}+\omega _{1}=0$ to solve (\ref%
{zwei}), stressing at this point that the constant of integration $\mu _{1}$
might be complex. At order $\lambda ^{2}$ we need to solve 
\begin{eqnarray}
\left( D_{x}^{4}+D_{x}D_{t}\right) (\tau ^{1}\cdot \tau ^{1}) &=&-\left(
D_{x}^{4}+D_{x}D_{t}\right) (1\cdot \tau ^{2}+\tau ^{2}\cdot 1)  \label{l2}
\\
&=&-2(\tau ^{2})_{xt}-2(\tau ^{2})_{xxxx}.
\end{eqnarray}%
Using $D_{x}^{m}D_{t}^{n}(e^{k_{i}x+\omega _{i}t+\mu _{i}}\cdot
e^{k_{j}x+\omega _{j}t+\mu _{j}})=(k_{i}-k_{j})^{m}(\omega _{i}-\omega
_{j})^{n}e^{k_{i}x+\omega _{i}t+\mu _{i}}e^{k_{j}x+\omega _{j}t+\mu _{j}}$
for $n,m\in \mathbb{N}_{0}$ this is easily achieved by setting $\tau ^{2}=0$%
. Then all higher order terms vanish by setting $\tau ^{k}=0$ for $k>2$.
Thus an exact $\tau $-function and corresponding solution to the KdV
equation are simply%
\begin{equation}
\tau _{\mu ,\beta }(x,t)=1+e^{\beta x-\beta ^{3}t+\mu },\quad \text{and\quad 
}u_{\mu ,\beta }(x,t)=\frac{\beta ^{2}}{2}\func{sech}\left[ \frac{1}{2}%
(\beta x-\beta ^{3}t+\mu )\right] ^{2},  \label{tau}
\end{equation}%
where we have set $\omega _{1}=-\beta ^{3}$, $k_{1}=\beta $, $\mu _{1}=\mu $
in order to satisfy the dispersion relation and $\lambda =1$. The standard
choices are here $\mu =0$ and $\mu =i\pi $ giving rise to the well-known
real solutions 
\begin{equation}
u_{0,\beta }(x,t)=\frac{\beta ^{2}}{2}\func{sech}\left[ \frac{1}{2}(\beta
x-\beta ^{3}t)\right] ^{2}\quad \text{and\quad }u_{i\pi ,\beta }(x,t)=\frac{%
\beta ^{2}}{2}\func{csch}\left[ \frac{1}{2}(\beta x-\beta ^{3}t)\right] ^{2}%
\text{,}
\end{equation}%
that may also be obtained from direct integration of the KdV equation with
appropriate boundary condition assuming the solutions to be travelling
waves. However, it is clear that any choice for which $\mu $ is purely
imaginary, i.e. $\mu =i\theta $ with $\theta \in \mathbb{R}$, would
constitute a finite $\mathcal{PT}$-invariant solution. Separating this
solution into its real and imaginary part we obtain%
\begin{equation}
u_{i\theta ,\beta }(x,t)=\frac{\beta ^{2}+\beta ^{2}\cos \theta \cosh (\beta
x-\beta ^{3}t)}{\left[ \cos \theta +\cosh (\beta x-\beta ^{3}t)\right] ^{2}}%
-i\frac{\beta ^{2}\sin \theta \sinh (\beta x-\beta ^{3}t)}{\left[ \cos
\theta +\cosh (\beta x-\beta ^{3}t)\right] ^{2}}.  \label{newS}
\end{equation}%
This form also allows explicitly to identify the solutions to the coupled
equation (\ref{KdV}) by just reading off the real and imaginary parts. For
the choice $\theta =\pm \pi /2$ this solution reduces precisely to the one
found by Khare and Saxena in \cite{khare2015novel}, up to an overall minus
sign due to the difference in (\ref{KdV}). We notice that while the $%
\mathcal{PT}$-invariance of $u_{\mu ;\beta }(x,t)$ is apparent, the one for
the corresponding $\tau $-functions $\tau _{\mu ;\beta }(x,t)$ are not
immediately obvious, in fact they are not $\mathcal{PT}$-invariant. This is
due to the ambiguity in those functions, as for instance $\tau
(x,t)\rightarrow u_{1}(x,t)\exp \left[ c_{1}x+c_{2}+f(t)\right] $ with
arbitrary constants $c_{1}$, $c_{2}$ and function $f(t)$ will give rise to
the same solution $u(x,t)$ to the KdV equation. Instead of taking the
standard form in (\ref{tau}) we may start from $\hat{\tau}_{\mu ;\beta
}(x,t)=\cosh \left[ \left( \beta x-\beta ^{3}t+\mu \right) /2\right] $
leading also to the same $u_{\mu ;\beta }(x,t)$ in (\ref{tau}). In this form
the $\mathcal{PT}$-invariance is directly evident. In other words the $\tau $%
-functions do not need to be $\mathcal{PT}$-symmetric in order to generate a 
$\mathcal{PT}$-symmetric solution for the KdV equation.

Let us next construct a two-soliton solution. As a starting point we take%
\begin{equation}
\tau ^{1}=e^{\eta _{1}}+e^{\eta _{2}},  \label{t1}
\end{equation}%
which naturally solves (\ref{zwei}) with nonlinear dispersion relations $%
k_{i}^{3}+\omega _{i}=0$ for $i=1,2$. At order $\lambda ^{2}$ we determine
from (\ref{l2}) that 
\begin{equation}
\tau ^{2}=\gamma e^{\eta _{1}+\eta _{2}}.  \label{t2}
\end{equation}%
with $\gamma =(\alpha -\beta )^{2}/(\alpha +\beta )^{2}$. The equation
resulting at order $\lambda ^{3}$ 
\begin{equation}
\left( D_{x}^{4}+D_{x}D_{t}\right) (1\cdot \tau ^{3}+\tau ^{1}\cdot \tau
^{2}+\tau ^{2}\cdot \tau ^{1}+\tau ^{3}\cdot 1)=0,
\end{equation}%
is solved by $\tau ^{1}$ and $\tau ^{2}$ given in (\ref{t1}) and (\ref{t2})
when setting $\tau ^{3}=0$. Once again all higher order equations are also
satisfied when setting $\tau ^{k}=0$ for $k\geq 3$, so that with $\lambda =1$%
\begin{equation}
\tau =1+e^{\eta _{1}}+e^{\eta _{2}}+\gamma e^{\eta _{1}+\eta _{2}},
\end{equation}%
becomes an exact solution to the Hirota equation (\ref{Hirota}), with $\mu
_{i}$ as defined in (\ref{etac}) possibly being complex. Translating the $%
\tau $-function back to the $u$-variable we obtain the two-soliton solution 
\begin{eqnarray}
&&u_{\mu ,\nu ;\alpha ,\beta }^{H}(x,t)=\frac{2\left( \beta ^{2}e^{2t\alpha
^{3}+t\beta ^{3}+x\beta +\mu }+\alpha ^{2}e^{t\alpha ^{3}+2t\beta
^{3}+\alpha x+\nu }\right) }{\left[ e^{t\alpha ^{3}+t\beta ^{3}}+e^{t\alpha
^{3}+x\beta +\mu }+e^{t\alpha ^{3}+x\beta +\nu }+\gamma e^{\mu +\nu +x\alpha
+x\beta )}\right] ^{2}}  \label{H2ss} \\
&&~~~~~~~+\frac{2\gamma e^{\mu +\nu }\left( 2(\alpha +\beta )^{2}e^{t\alpha
^{3}+t\beta ^{3}+x\alpha +x\beta }+\alpha ^{2}e^{\mu +\alpha ^{3}t+\alpha
x+2\beta x}+\beta ^{2}e^{\nu +\beta ^{3}t+2\alpha x+\beta x}\right) }{\left[
e^{t\alpha ^{3}+t\beta ^{3}}+e^{t\alpha ^{3}+x\beta +\mu }+e^{t\alpha
^{3}+x\beta +\nu }+\gamma e^{\mu +\nu +x\alpha +x\beta )}\right] ^{2}}. 
\notag
\end{eqnarray}%
Notice that (\ref{H2ss}) is not $\mathcal{PT}$-symmetric, even for the real
solution when taking $\mu =\nu =0$. It is evident that further multi-soliton
solutions constructed by means of the Hirato method will also not be $%
\mathcal{PT}$-symmetric. However, as we will show in section 2.2 that does
not mean that all multi-soliton solutions have broken $\mathcal{PT}$%
-symmetry. Moreover, it will turn out that despite having broken $\mathcal{PT%
}$-symmetry their corresponding energies are real. In the next subsection we
shall demonstrate that $\mathcal{PT}$-symmetric multi-soliton solutions may
be constructed from B\"{a}cklund transformations instead.

\subsubsection{Complex solutions from B\"{a}cklund transformations}

Converting the KdV equation (\ref{KdV}) into an equation for the quantity $w$%
, defined via $u=w_{x}$, the KdV-B\"{a}cklund transformations are well known
to relate two different solutions $u,w$ and $u^{\prime },w^{\prime }$ as%
\begin{eqnarray}
w_{x}+w_{x}^{\prime } &=&\kappa -\frac{1}{2}(w-w^{\prime })^{2},
\label{kdvb1} \\
w_{t}+w_{t}^{\prime } &=&(w-w^{\prime })(w_{xx}-w_{xx}^{\prime
})-2[w_{x}^{2}+w_{x}w_{x}^{\prime }+(w_{x}^{\prime })^{2}].  \label{kdvb2}
\end{eqnarray}%
A \textquotedblleft nonlinear superposition principle\textquotedblright\ is
then obtained by relating four different solutions as $w_{0}\overset{\kappa
_{1}}{\rightarrow }w_{1}$, $w_{0}\overset{\kappa _{2}}{\rightarrow }w_{2}$, $%
w_{1}\overset{\kappa _{2}}{\rightarrow }w_{12}$ and $w_{2}\overset{\kappa
_{1}}{\rightarrow }w_{12}$. Using the corresponding four versions of (\ref%
{kdvb1}) all differentials may be eliminated, such that one can construct a
new solution $w_{12}$ to the KdV equation from three known solutions $w_{0}$%
, $w_{1}$ and $w_{2}$ as 
\begin{equation}
w_{12}=w_{0}+2\frac{\kappa _{1}-\kappa _{2}}{w_{1}-w_{2}}.
\end{equation}%
With $w_{\mu ;\beta }(x,t)=\beta \tanh \left[ \frac{1}{2}(\beta x-\beta
^{3}t+\mu )\right] $, resulting from $u_{\mu ;\beta }(x,t)$ in (\ref{tau}),
we identify $\kappa =\beta ^{2}/2$ from (\ref{kdvb1}) when taking $%
w=w_{0;\beta }=0$ and $w^{\prime }=w_{\mu ;\beta }$. A new solution to the
KdV equation is therefore%
\begin{equation}
w_{\mu ,\nu ;\alpha ,\beta }=\frac{\alpha ^{2}-\beta ^{2}}{w_{\mu ;\alpha
}-w_{\nu ;\beta }},
\end{equation}%
with corresponding wavefunction%
\begin{equation}
u_{\mu ,\nu ;\alpha ,\beta }^{B}(x,t)=\frac{\alpha ^{2}-\beta ^{2}}{2}\frac{%
\beta ^{2}\func{sech}\left[ \frac{1}{2}(\beta x-\beta ^{3}t+\nu )\right]
^{2}-\alpha ^{2}\func{sech}\left[ \frac{1}{2}(\alpha x-\alpha ^{3}t+\mu )%
\right] ^{2}}{\left[ \alpha \tanh \left[ \frac{1}{2}(\alpha x-\alpha
^{3}t+\mu )\right] -\beta \tanh \left[ \frac{1}{2}(\beta x-\beta ^{3}t+\nu )%
\right] \right] ^{2}}.  \label{twos}
\end{equation}

\FIGURE{ \epsfig{file=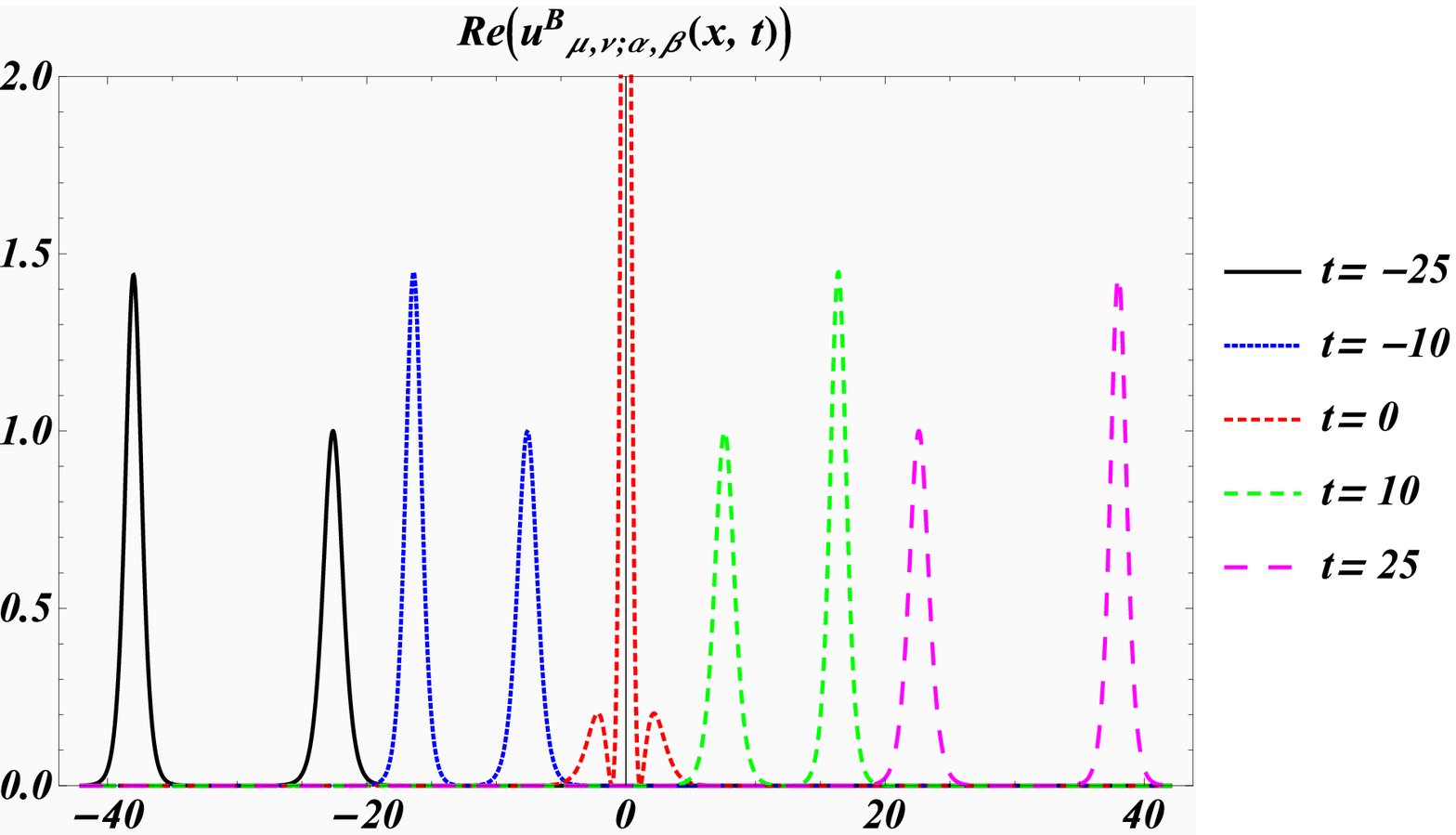,height=4.6cm} \epsfig{file=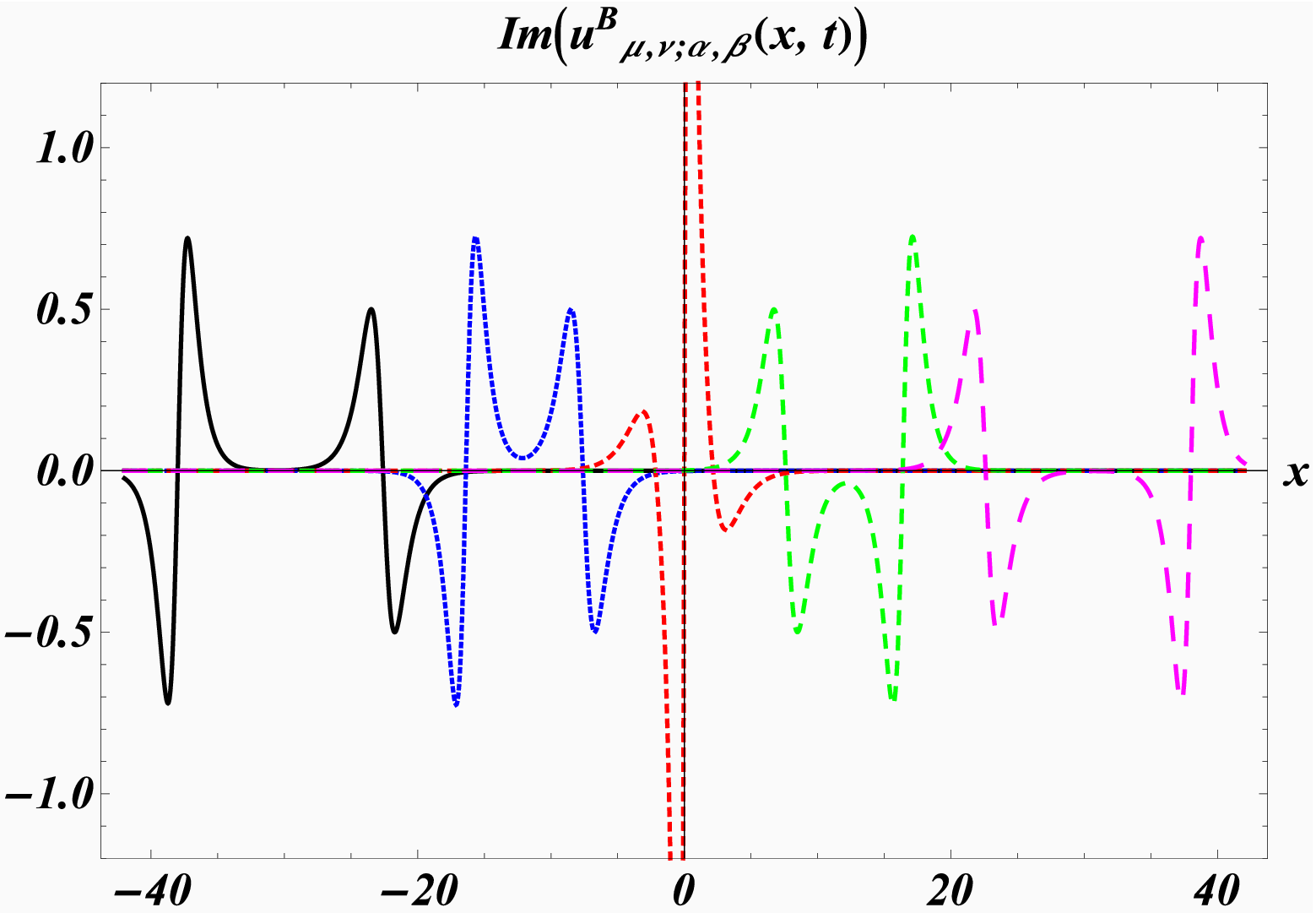,height=4.6cm}
\caption{$\mathcal{PT}$-symmetric KdV two-soliton solution with $\alpha=6/5$, $\beta=1$ and
$\mu=\nu=i \pi/2$.}
        \label{KdV2PT}}

Notice that, unlike for $u_{\mu ,\nu ;\alpha ,\beta }^{H}$, for real values
of $\mu $ and $\nu $ the denominator of $u_{\mu ,\nu ;\alpha ,\beta }^{B}$
vanishes at certain values for $x$ and $t$. Thus complex values for $\mu $
and $\nu $ can be used to regularize this expression. We observe further
that while the $\mathcal{PT}$-symmetric one-soliton solutions may formally
be obtained simply from complex shifts in space or time from one basic
solution $u_{0;\beta }(x,t)$, neither the broken $\mathcal{PT}$-symmetric
two-soliton $u^{H}$ nor the $\mathcal{PT}$-symmetric two-soliton $u^{B}$ is
obtainable from a known two-soliton solution in this simple manner when $\mu
\neq \nu $. However, we may use real shifts in space or time to restore the $%
\mathcal{PT}$-symmetry for the broken $\mathcal{PT}$-symmetric one-soliton
solution $u_{\mu _{r}+i\mu _{i}}$, with $\mu _{r},\mu _{i}\in \mathbb{R}$,
as 
\begin{equation}
u_{\mu _{r}+i\mu _{i};\beta }\left( x-\frac{\mu _{r}}{\beta },t\right)
=u_{\mu _{r}+i\mu _{i};\beta }\left( x,t+\frac{\mu _{r}}{\beta ^{3}}\right)
=u_{i\mu _{i};\beta }\left( x,t\right) .  \label{tsshift}
\end{equation}%
To achieve this restoration for the broken $\mathcal{PT}$-symmetric
two-soliton solution we require a simultaneous shift in space and time 
\begin{equation}
u_{\mu _{r}+i\mu _{i},\nu _{r}+i\nu _{i};\alpha ,\beta }^{B}\left( x+\frac{%
\beta ^{3}\mu _{r}-\alpha ^{3}\nu _{r}}{\alpha ^{3}\beta -\alpha \beta ^{3}}%
,t+\frac{\beta \mu _{r}-\alpha \nu _{r}}{\alpha ^{3}\beta -\alpha \beta ^{3}}%
\right) =u_{i\mu _{i},i\nu _{i};\alpha ,\beta }^{B}\left( x,t\right) .
\label{sh}
\end{equation}

In figure \ref{KdV2PT} we display the two-soliton solution $u_{\mu ,\nu
;\alpha ,\beta }^{B}$ for a $\mathcal{PT}$-symmetric choice of the
parameters $\mu =\nu $. We observe $\func{Re}[u^{B}(x,t)]=\func{Re}%
[u^{B}(-x,-t)]$ and also $\func{Im}[u^{B}(x,t)]=-\func{Im}[u^{B}(-x,-t)]$.
The real part exhibits the typical features of a two-soliton scattering,
that is being separated into two one-soliton solutions in the past and
regaining the original shapes with exchanged positions in the future, with a
time-delay as the only residual effect. For the complex solutions this
behaviour is now accompanied by a smooth scattering structure for the
imaginary part. In as similar fashion as in \cite%
{rubinstein,jackiw,fring1994vertex} we compute the time-delay for the real
and imaginary parts as%
\begin{eqnarray}
\lim_{t\rightarrow \pm \infty }\func{Re}\left[ u_{i\pi /2,i\pi /2;\alpha
,\beta }^{B}\left( x,t\right) \right] &=&\func{Re}\left[ u_{i\pi /2;\beta
}^{B}\left( x,t\pm \Delta _{\beta }\right) \right] +\func{Re}\left[ u_{i\pi
/2;\alpha }^{B}\left( x,t\mp \Delta _{\alpha }\right) \right] ,~~\ \ \ \ 
\label{treal} \\
\lim_{t\rightarrow \pm \infty }\func{Im}\left[ u_{i\pi /2,i\pi /2;\alpha
,\beta }^{B}\left( x,t\right) \right] &=&\func{Im}\left[ u_{i\pi /2;\beta
}^{B}\left( x,t\pm \Delta _{\beta }\right) \right] -\func{Im}\left[ u_{i\pi
/2;\alpha }^{B}\left( x,t\mp \Delta _{\alpha }\right) \right] ,  \label{tim}
\end{eqnarray}%
where the time shifts are given by%
\begin{equation}
\Delta _{x}=\frac{1}{x^{3}}\ln \left[ \frac{\alpha +\beta }{\alpha -\beta }%
\right] .  \label{shift}
\end{equation}%
We confirm our analytic results by numerical computations displayed in
figure \ref{Retime}.

\FIGURE{ \epsfig{file=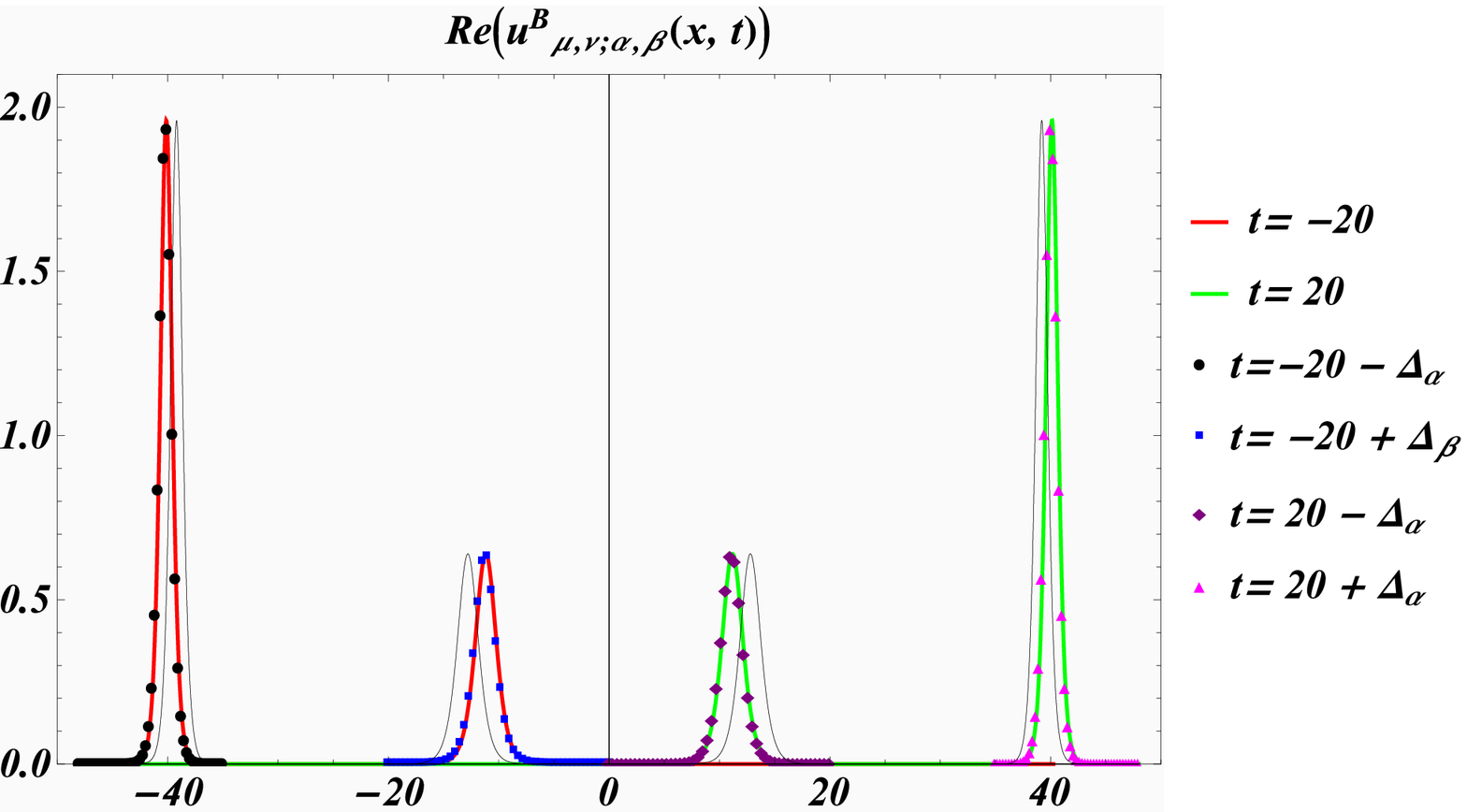,height=4.5cm} \epsfig{file=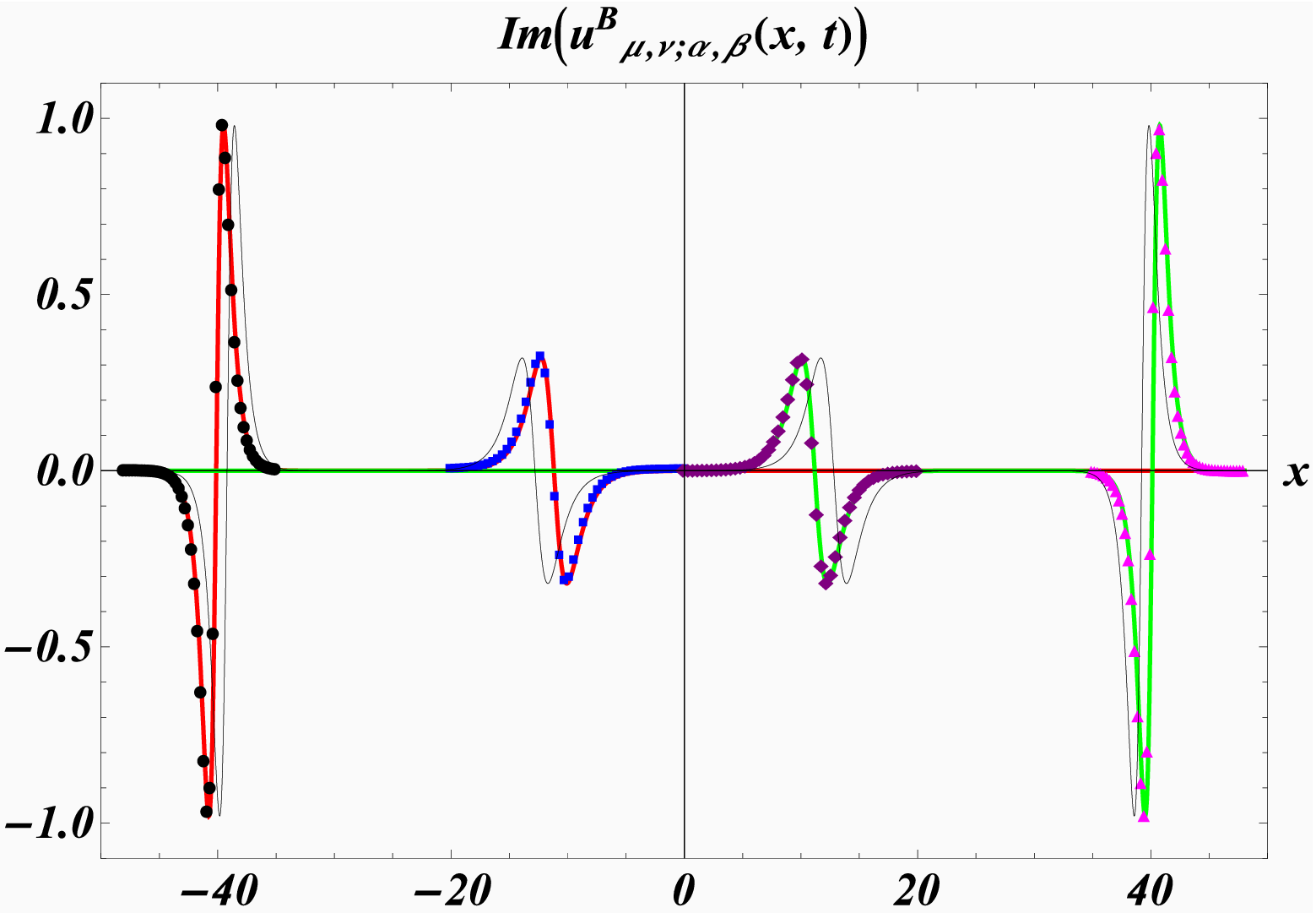,height=4.5cm}
        \caption{Time delay in the KdV complex two-soliton scattering $u_{\mu,\nu ;\alpha
,\beta }^{B}$ for $\mu=\nu=i\pi /2$, $\alpha=7/5$ and $\beta=4/5$. The scattered solitons are the 
one-solition solutions at the shifted times as indicated in the legend and the thin solid black
lines are the unshifted one-soliton solutions at time $t=-20,-20,20,20$ from the left to the right.}
        \label{Retime}}

We observe a perfect match between the two-soliton solutions and the $\Delta
_{x}$-shifted one-soliton solution in the real as well as in the imaginary
part. The faster soliton, i.e. the one related to $\alpha $ in our choice of
parameters, in the two-soliton solution is shifted to the left in the past
and to the right in the future. These shifts are in the opposite direction
for the slower soliton related to $\beta $. The details of the derivation
for (\ref{treal}), (\ref{tim}), (\ref{shift}) together with a some further
analysis are presented elsewhere \cite{timedelay}.

As seen in figure \ref{KdV2PTbroken} the qualitative behaviour does not
change in the broken regime, with the only difference that two solutions for
some specific values $t^{\prime }$ and $-t^{\prime }$ are no longer
symmetric around $x=0$, similarly as for the solution $u^{B}$. Taking $\mu $
different from $\nu $ we can modulate the shapes of the different solutions
as displayed in figure \ref{KdV2PTDiff}.

\FIGURE{ \epsfig{file=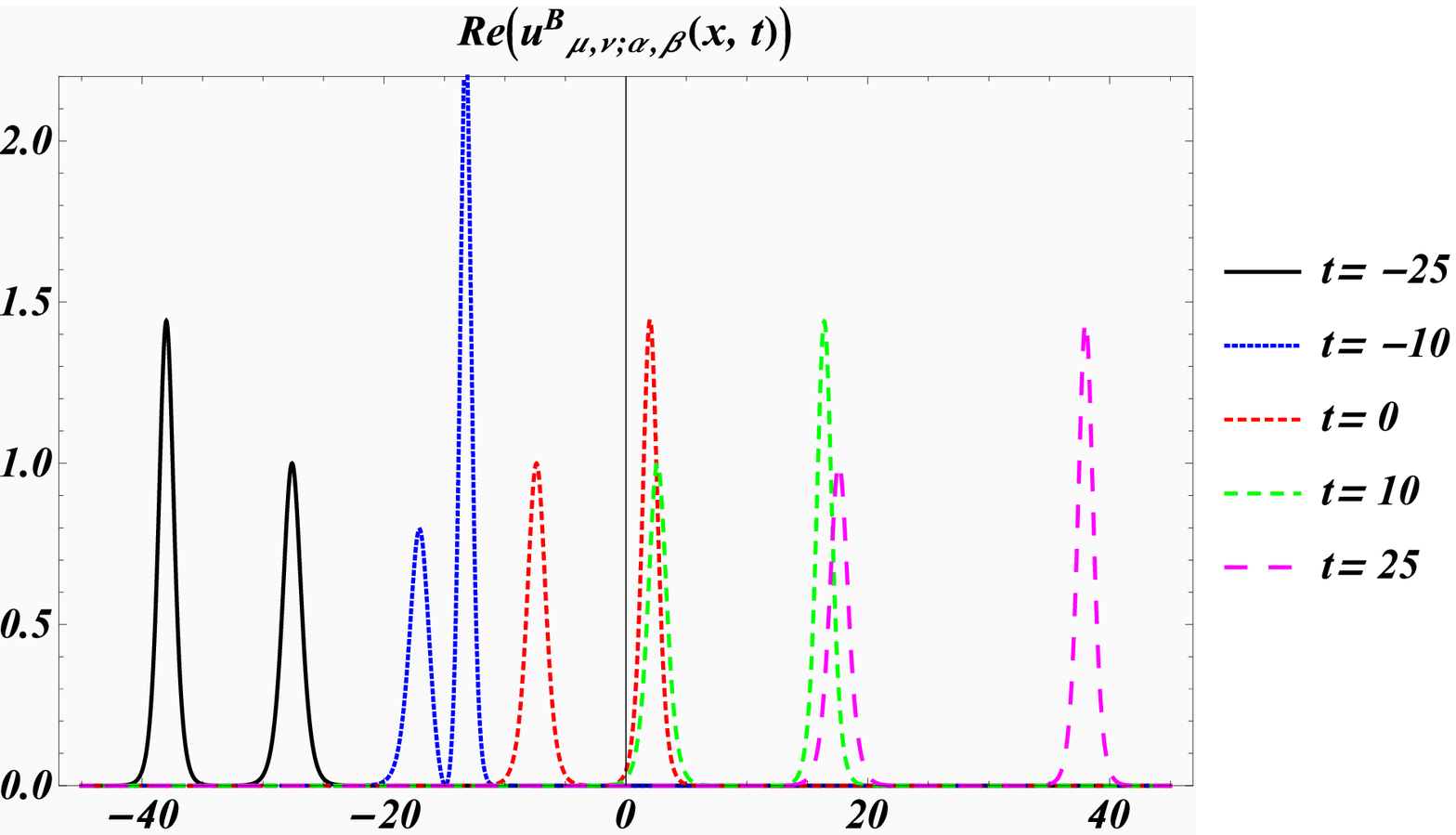,height=4.6cm} \epsfig{file=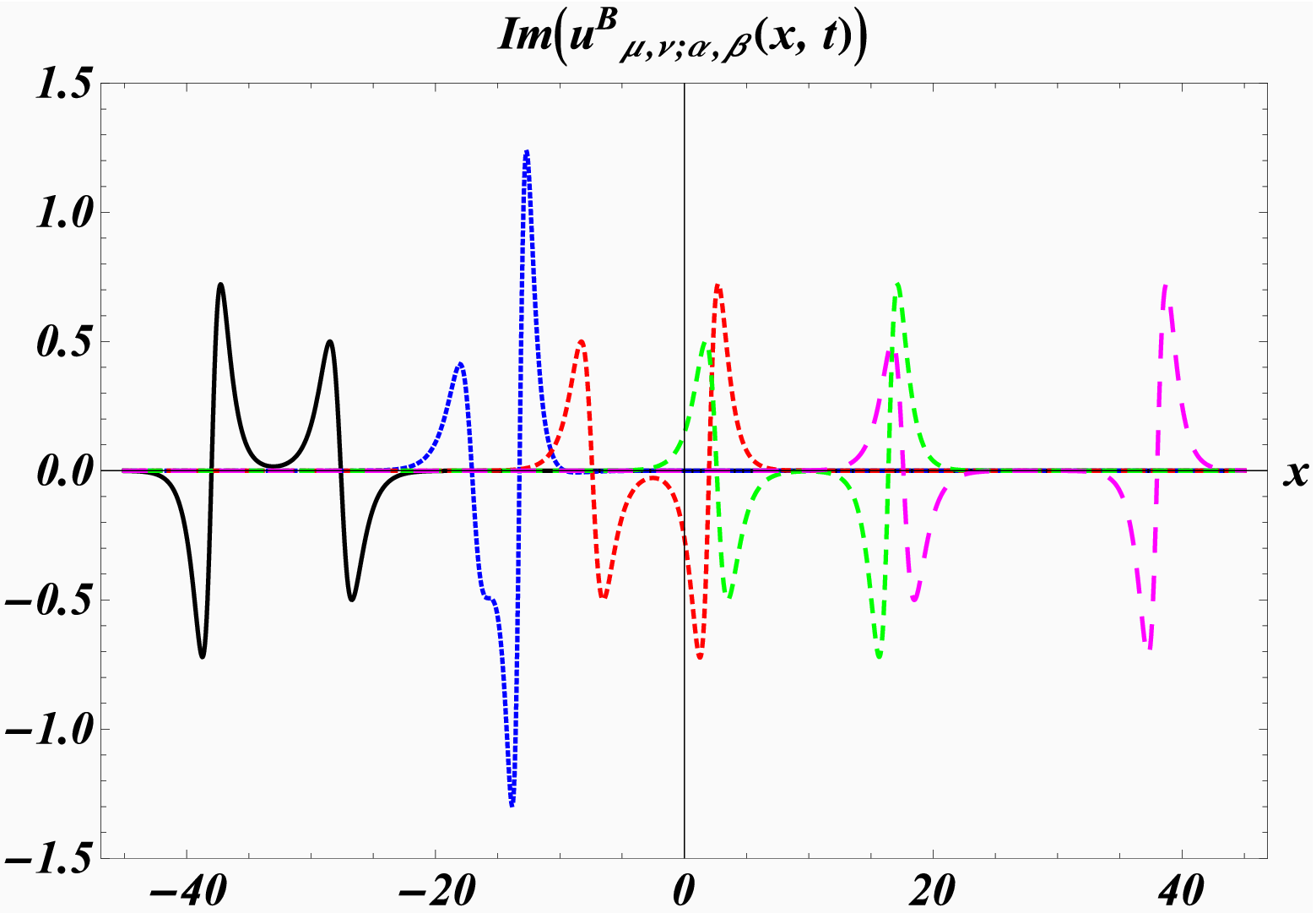,height=4.6cm}
        \caption{Broken $\mathcal{PT}$-symmetric KdV two-soliton solution with $\alpha=6/5$, 
$\beta=1$, $\mu=5+ i \pi/2$ and $\nu=i \pi/2$.}
        \label{KdV2PTbroken}}

\FIGURE{ \epsfig{file=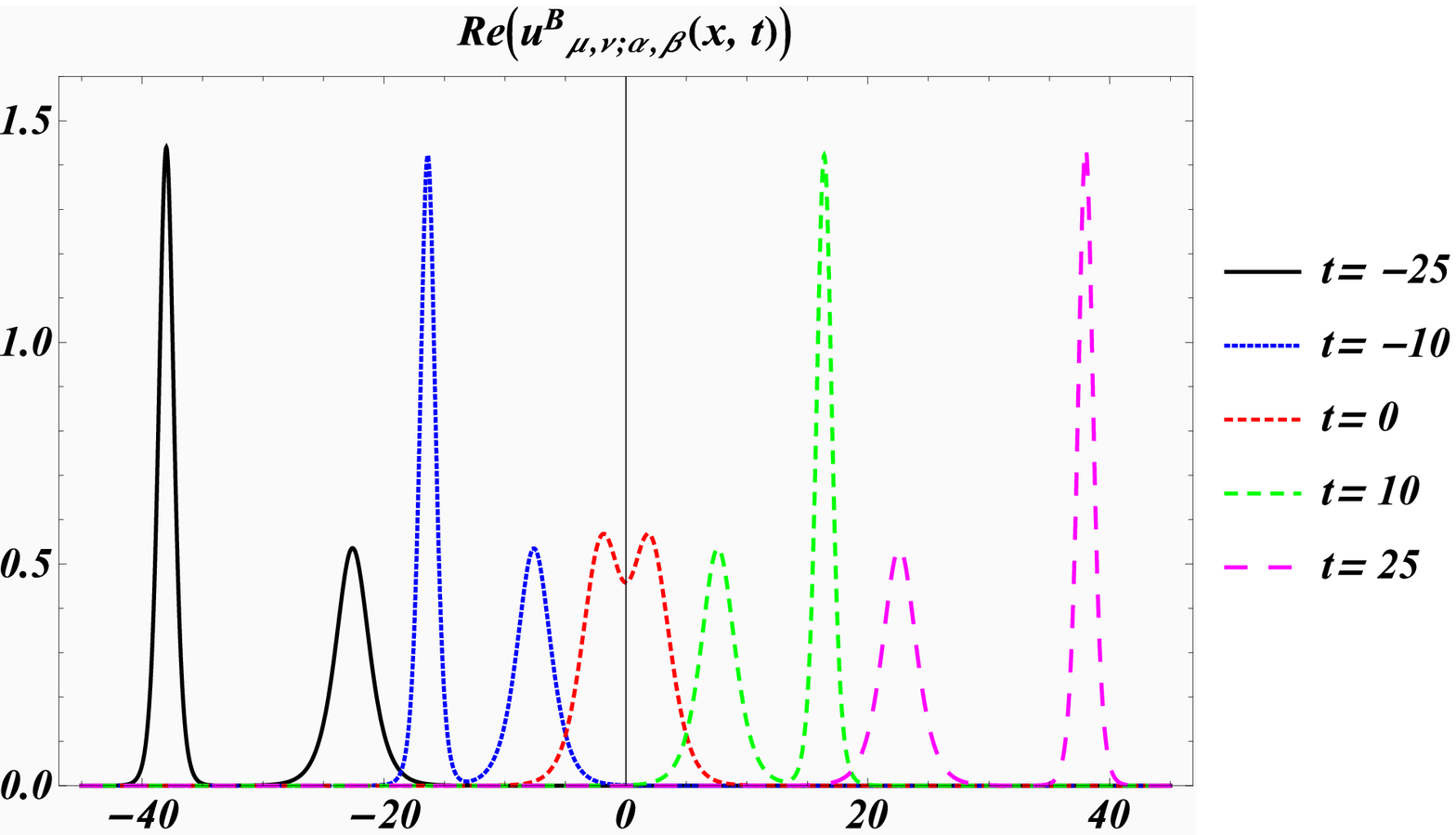,height=4.6cm} \epsfig{file=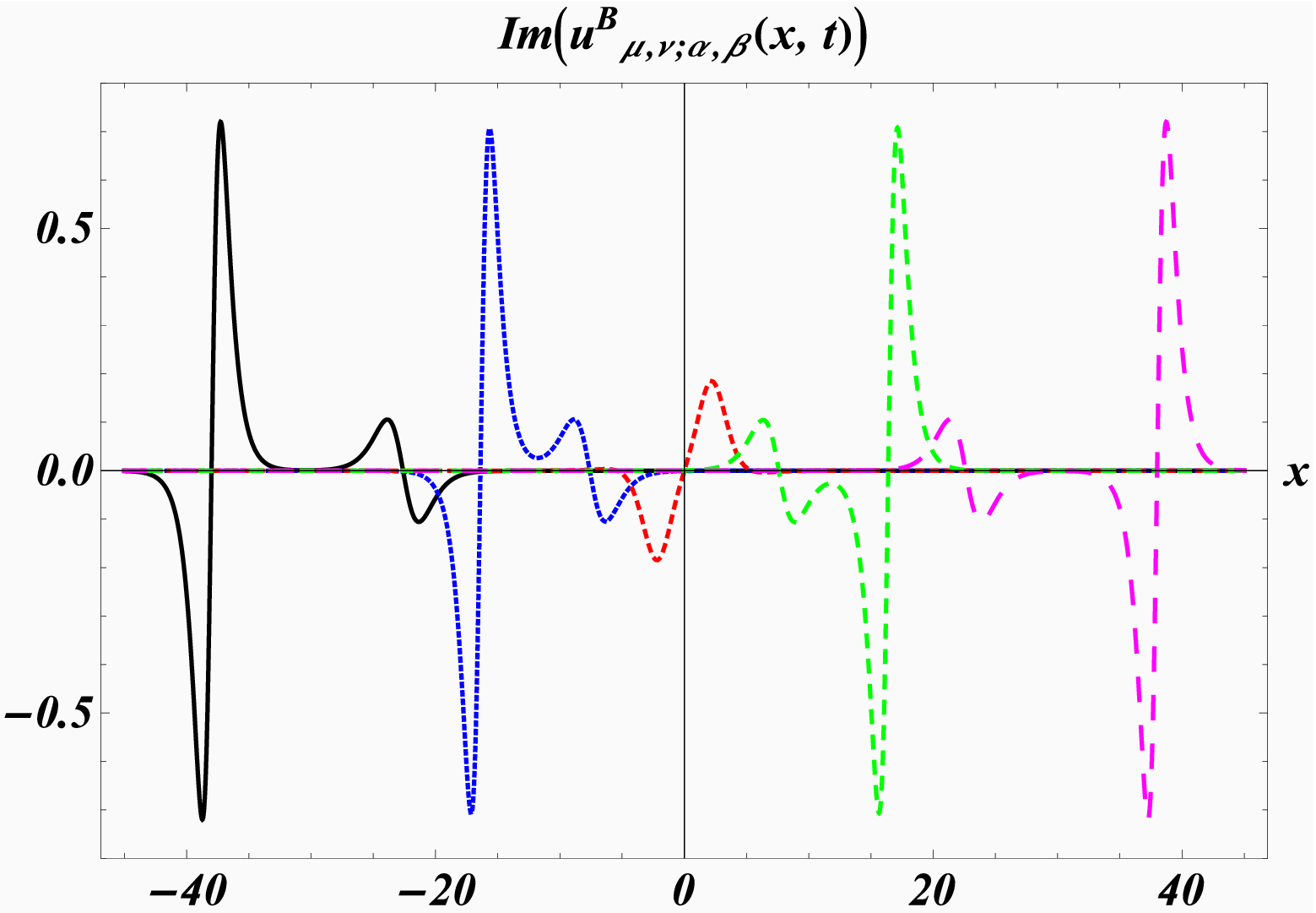,height=4.6cm}
        \caption{$\mathcal{PT}$-symmetric KdV two-soliton solution from different types of one-solitons with $\alpha=6/5$, $\beta=4/5$, $\mu=i\pi/6$ and $\nu=i \pi/2$.}
        \label{KdV2PTDiff}}

\subsubsection{Real energies from $\mathcal{PT}$-symmetric and broken $%
\mathcal{PT}$-symmetric solutions}

Having obtained various types of solutions, we will now compute the
corresponding energies resulting from the expression (\ref{E}). The
Hamiltonian density leading to the KdV equation in the form (\ref{KdV}) is
given by%
\begin{equation}
\mathcal{H}(u,u_{x})=-u^{3}+\frac{1}{2}u_{x}^{2}.
\end{equation}%
From the KdV B\"{a}cklund transformation (\ref{kdvb1}) with $w^{\prime }=0$
and the observation that $(u_{\mu ;\beta })_{x}/u_{\mu ;\beta }=-w_{\mu
;\beta }=-\beta \tanh \left[ \frac{1}{2}(\beta x-\beta ^{3}t+\mu )\right] $,
we derive the identity $(u_{\mu ;\beta })_{xx}=$ $(u_{\mu ;\beta
})_{x}^{2}/u_{\mu ;\beta }-u_{\mu ;\beta }^{2}$. These relations allow us to
write the Hamiltonian density as%
\begin{equation}
\mathcal{H}\left[ u_{\mu ;\beta },(u_{\mu ;\beta })_{x}\right] =\left\{ 
\frac{1}{10}\left[ \frac{(u_{\mu ;\beta })_{x}}{u_{\mu ;\beta }}\right] ^{5}+%
\frac{1}{2}\left[ \frac{(u_{\mu ;\beta })_{x}}{u_{\mu ;\beta }}\right]
^{2}(u_{\mu ;\beta })_{x}+u_{\mu ;\beta }(u_{\mu ;\beta })_{x}\right\} _{x}.
\end{equation}%
The corresponding energy then simply results to%
\begin{equation}
E_{\mu ;\beta }=\int\nolimits_{-\infty }^{\infty }{\mathcal{H}}[u_{\mu
;\beta },(u_{\mu ;\beta })_{x}]dx=\frac{1}{2}\left[ \frac{1}{5}\left[ \frac{%
(u_{\mu ;\beta })_{x}}{u_{\mu ;\beta }}\right] ^{5}+\frac{(u_{\mu ;\beta
})_{x}^{3}}{u_{\mu ;\beta }^{2}}+(u_{\mu ;\beta }^{2})_{x}\right] _{-\infty
}^{\infty }=-\frac{\beta ^{5}}{5},  \label{Emu}
\end{equation}%
when using asymptotically vanishing boundary conditions for the wave
function and its derivative $\lim_{x\rightarrow \pm \infty }u_{\mu ;\beta
}=\lim_{x\rightarrow \pm \infty }(u_{\mu ;\beta })_{x}=0$ together with $%
\lim_{x\rightarrow \pm \infty }$ $(u_{\mu ;\beta })_{x}/u_{\mu ;\beta }=\mp
\beta $. Notice that as long as $\beta $ is real this energy is real at all
times $t$, irrespective of whether $u_{\mu ;\beta }$ is $\mathcal{PT}$%
-symmetric or not. The reason is simple: Taking $\mu $ to be of the form $%
\mu =\mu _{r}+i\mu _{i}$, the $\mathcal{PT}$-symmetry of ${\mathcal{H}}$ is
broken when $\mu _{r}\neq 0$. However, a simple shift in time or space, as
explained in (\ref{tsshift}), will restore the $\mathcal{PT}$-symmetry of
the integrand. Both type of shifts are permitted, as the shift in $x$ can be
absorbed in the limits of the integral and the shift in $t$ is allowed since 
$\mathcal{H}$ is a conserved quantity in time. As argued before, having a $%
\mathcal{PT}$-symmetric integrand the complex part does not contribute to
the overall value of $E_{\mu ;\beta }$.

For the two-soliton solutions $u_{\mu ,\nu ;\alpha ,\beta }^{H,B}(x,t)$, we
compute numerically that the total energy is the sum of the individual
one-soliton solutions%
\begin{equation}
E_{\mu ,\nu }^{H,B}=\int\nolimits_{-\infty }^{\infty }{\mathcal{H}}\left[
u_{\mu ,\nu ;\alpha ,\beta }^{H,B}(x,t),\left( u_{\mu ,\nu ;\alpha ,\beta
}^{H,B}(x,t)\right) _{x}\right] dx=E_{\mu ;\beta }+E_{\nu ;\alpha }=-\frac{%
\alpha ^{5}+\beta ^{5}}{5}.  \label{HB}
\end{equation}%
Once again we notice that we obtain real energies also for the $\mathcal{PT}$%
-symmetrically broken scenario. In this case we can restore the $\mathcal{PT}
$-symmetry by a simultaneous shift in $x$ and $t$ as explained in (\ref{sh}%
). While this explains the reality of the spectrum, it does not yet account
for the concrete values in (\ref{HB}). However, as we have seen in (\ref%
{treal}) and (\ref{tim}) for one specific case, asymptotically the
two-soliton solution separates into two from each other isolated one-soliton
solutions, in both the real and imaginary part. These one-soliton solutions
contribute separately to the total energy, which is the same value at all
times. As the latter argument applies to any $N$-soliton solution we expect
their energies to be the sum of all their $N$ asymptotic individual
one-soliton solutions. However this still needs verification \cite{timedelay}%
.

\subsection{The complex modified Korteweg-de Vries equation}

Using the variable transformation $v=\hat{w}_{x}$ the mKdV equation can be
written in the two equivalent forms 
\begin{equation}
v_{t}+24v^{2}v_{x}+v_{xxx}=0\quad \Leftrightarrow \quad \hat{w}_{t}+8\hat{w}%
_{x}^{3}+\hat{w}_{xxx}=0.  \label{mKdV}
\end{equation}%
Unlike the KdV equation, the mKdV equation allows for two alternative types
of $\mathcal{PT}$-symmetries $\mathcal{PT}_{\pm }$: $x\rightarrow -x$,\ $%
t\rightarrow -t$, $i\rightarrow -i$, $v\rightarrow \pm v$. With the further
substitution $\hat{w}=\arctan (\tau /\sigma )$ the latter equation in (\ref%
{mKdV}) can be converted into Hirota's bilinear form \cite{hirotamkdv} 
\begin{equation}
\left( D_{t}+D_{x}^{3}\right) \tau \cdot \sigma =0,\qquad \text{and\qquad }%
D_{x}^{2}\left( \tau \cdot \tau +\sigma \cdot \sigma \right) =0,  \label{DD}
\end{equation}%
when using the relations (\ref{H1})-(\ref{H5}). Taking now $\sigma =1$ the
equations (\ref{DD}) reduce to 
\begin{equation}
\tau _{t}+\tau _{xxx}=0,\qquad \text{and\qquad }\tau \tau _{xx}-\tau
_{x}^{2}=0.
\end{equation}%
The exact solutions to these equations with corresponding solution to the
mKdV equation (\ref{mKdV}) are 
\begin{equation}
\tau _{\mu ;\beta }(x,t)=e^{\beta x-\beta ^{3}t+\mu },\quad \text{and\quad }%
v_{\mu ;\beta }(x,t)=\frac{\beta }{2}\func{sech}\left[ \beta x-\beta
^{3}t+\mu \right] .  \label{smkdv}
\end{equation}%
It is well known that the mKdV and the KdV equation are related by a Miura
transformation. Here we find that the solutions (\ref{tau}) and (\ref{smkdv}%
) to the KdV equations and mKdV equation (\ref{KdV}) and (\ref{mKdV}),
respectively, are related as 
\begin{equation}
u_{\mu \pm i\frac{\pi }{2};\beta }(x,t)=4v_{\mu ;\beta }^{2}(x,t)\pm
i2\partial _{x}v_{\mu ;\beta }(x,t).  \label{Miura}
\end{equation}%
This means for instance that the real solution $v_{0;\beta }(x,t)$ to the
mKdV equation leads inevitably to the complex $\mathcal{PT}$-symmetric
solutions $u_{\pm i\frac{\pi }{2};\beta }(x,t)$ for the KdV equation. Thus
we have obtained yet another way to derive the solutions reported in \cite%
{khare2015novel}. The complex part simply results from scaling the more
familiar transformation $u=v^{2}+v_{x}$, that relates the mKdV with
nonlinear term $-6v^{2}v_{x}$ to the KdV equation with nonlinear term $%
+6uu_{x}$, to the present forms (\ref{KdV}) and (\ref{mKdV}).

The latter argument may also be applied to solutions in terms of Jacobi
elliptic functions. Starting with the shifted known solution to the mKdV
equation 
\begin{equation}
\hat{v}_{\mu ;\beta }(x,t)=\frac{\beta }{2}\func{dn}\left[ \beta x-\beta
^{3}t(2-m)+\mu ,m\right] ,  \label{el1}
\end{equation}%
we obtain from (\ref{Miura}) the corresponding solution to the KdV equation%
\begin{equation}
\hat{u}_{\mu ;\beta }^{\pm }(x,t)=\beta ^{2}\func{dn}\left[ \hat{z},m\right]
^{2}\pm im\beta ^{2}\func{cn}\left[ \hat{z},m\right] \func{sn}\left[ \hat{z}%
,m\right] ,
\end{equation}%
where we abbreviated the argument $\hat{z}:=\beta x-\beta ^{3}t(2-m)+\mu $.
The elliptic parameter is denoted by $m$ as usual. Likewise from the shifted
known solution to the mKdV equation%
\begin{equation}
\tilde{v}_{\mu ;\beta }(x,t)=\frac{\beta }{2}\sqrt{m}\func{cn}\left[ \beta
x-\beta ^{3}t(2m-1)+\mu ,m\right]  \label{el2}
\end{equation}%
we construct%
\begin{equation}
\tilde{u}_{\mu ;\beta }^{\pm }(x,t)=m\beta ^{2}\func{cn}\left[ \tilde{z},m%
\right] ^{2}\pm i\sqrt{m}\beta ^{2}\func{dn}\left[ \tilde{z},m\right] \func{%
sn}\left[ \tilde{z},m\right] ,
\end{equation}%
with $\tilde{z}:=\beta x-\beta ^{3}t(2m-1)+\mu $. Thus the solutions $\hat{v}%
_{0;\beta }(x,t)$ and $\tilde{v}_{0;\beta }(x,t)$ to the mKdV equation,
which could be real for specific values, lead to the complex $\mathcal{PT}$%
-symmetric solution for the KdV equation reported in \cite{khare2015novel}.
It is clear that this is only one possibility as other choices for purely
imaginary $\mu $ also respect the $\mathcal{PT}$-symmetry.

\subsubsection{Real energies from $\mathcal{PT}$-symmetric and broken $%
\mathcal{PT}$-symmetric solutions}

Next we compute the energy resulting from the mKdV Hamiltonian density
leading to the equation of motion (\ref{mKdV}) after variation%
\begin{equation}
\mathcal{H}(v,v_{x})=-2v^{4}+\frac{1}{2}v_{x}^{2}.
\end{equation}%
For the solution $v_{\mu ;\beta }$ in (\ref{smkdv}) we compute the energy%
\begin{equation}
E_{\mu ;\beta }=\int\nolimits_{-\infty }^{\infty }{\mathcal{H}}\left[ v_{\mu
;\beta }(x,t),\left( v_{\mu ;\beta }(x,t)\right) _{x}\right] dx=-\frac{\beta
^{3}}{12},
\end{equation}%
which has the same properties as the energy of the KdV one-soliton, that is
being real for all values of $\mu $. The elliptic solutions have the two
periods $4K(m)/\beta $ and $i4K(1-m)/\beta $ in $x$ with $K(m)$ denoting the
elliptic integral of the first kind. Thus we have to restrict the domain of
integration in (\ref{E}) in order to obtain finite energies. For the
solution $\hat{v}_{\mu ;\beta }$ in (\ref{smkdv}) we compute the real
energies%
\begin{eqnarray}
\hat{E}_{\mu ;\beta } &=&\int\nolimits_{-2K(m)/\beta }^{2K(m)/\beta }{%
\mathcal{H}}\left[ \hat{v}_{\mu ;\beta }(x,t),\left( \hat{v}_{\mu ;\beta
}(x,t)\right) _{x}\right] dx \\
&=&\frac{\beta ^{3}}{24}\left[ (m-2)E\left[ \limfunc{am}\left(
4K(m)|m\right) ,m\right] +4K(m)(m-1)\right] ,  \notag
\end{eqnarray}%
where $\limfunc{am}\left( u|m\right) $ denotes the amplitude of the Jacobi
elliptic function and $E\left[ \phi ,m\right] $ the elliptic integral of the
second kind. Similarly for the solution $\tilde{v}_{\mu ;\beta }$ in (\ref%
{el2}) we find 
\begin{eqnarray}
\tilde{E}_{\mu ;\beta } &=&\int\nolimits_{-2K(m)/\beta }^{2K(m)/\beta }{%
\mathcal{H}}\left[ \tilde{v}_{\mu ;\beta }(x,t),\left( \tilde{v}_{\mu ;\beta
}(x,t)\right) _{x}\right] dx \\
&=&\frac{\beta ^{3}}{24}\left[ (1-2m)E\left[ \limfunc{am}\left(
2K(m)|m\right) ,m\right] -4K(m)(3m^{2}-4m+1)\right] .  \notag
\end{eqnarray}%
We observe that $\lim_{m\rightarrow 1}\hat{E}_{\mu ;\beta
}=\lim_{m\rightarrow 1}\tilde{E}_{\mu ;\beta }=2E_{\mu ;\beta }$. For the
same reason as for the hyperbolic solutions all energies are real,
irrespective of whether the Hamiltonian densities are $\mathcal{PT}$%
-symmetric or not.

\subsection{The complex sine-Gordon equation}

The quantum field theory version of the complex sine-Gordon model has been
studied for some time \cite{CSG1,CSG2,demai,doholl,aratyn,okamura}. Here we
demonstrate that its classical version also admits interesting $\mathcal{PT}$%
-symmetric solutions with similar properties to those constructed in the
previous subsections. We consider the equation in the form 
\begin{equation}
\phi _{xt}=\sin \phi ,  \label{SG}
\end{equation}%
using light-cone variables, which we still call $x$ and $t$ with a slight
abuse of notation. We observe that this equation admits various symmetries
for $\mathcal{PT}_{\pm }^{(n)}$: $x\rightarrow -x$,\ $t\rightarrow -t$, $%
i\rightarrow -i$, $\phi \rightarrow \pm \phi +n2\pi $ with $n\in \mathbb{Z}$%
, with $\mathcal{PT}_{-}^{(n)}$ and $\mathcal{PT}_{+}^{(0)}$ squaring to 1
as expected for a proper $\mathcal{PT}$-symmetry. In \cite{HirotaSG} Hirota
showed that the sine-Gordon equation (\ref{SG}) can be converted into the
bilinear form 
\begin{equation}
D_{x}D_{t}\tau \cdot \sigma =\tau \cdot \sigma ,\qquad \text{and\qquad }%
D_{x}D_{t}\tau \cdot \tau =D_{x}D_{t}\sigma \cdot \sigma ,
\end{equation}%
when using the relations (\ref{H1})-(\ref{H5}) and the transformation $\phi
=4\arctan (\tau /\sigma )$. Taking $\sigma =1$ these equations reduce to 
\begin{equation}
\tau _{xt}=\tau \qquad \text{and\qquad }\tau \tau _{xt}=\tau _{x}\tau _{t}.
\end{equation}%
The exact solutions to these equations and therefore the corresponding
solutions to the sine-Gordon equation (\ref{SG}) are easily found. For
instance, we obtain the well-known kink solution as 
\begin{equation}
\tau _{\mu ;\beta }(x,t)=e^{\beta x+t/\beta +\mu },\quad \text{and\quad }%
\phi _{\mu ;\beta }(x,t)=4\arctan \left[ e^{\beta x+t/\beta +\mu }\right] .
\end{equation}%
Recalling that $\arctan z=-\arctan z^{-1}\pm \pi /2$ for $\func{Re}z\QATOP{>%
}{<}0$, we note that the solution for $\mu =i\theta $ with $\theta \in 
\mathbb{R}$ is $\mathcal{PT}_{-}^{(\pm )}$-symmetric. Let us separate off
the real and imaginary parts of the solution for these values of $\mu $ by
using the well-known relation $\arctan z=-i/2\ln \left[ (i-z)/(i+z)\right] $%
. For the principle value of the logarithm we obtain%
\begin{equation}
\phi _{i\theta ;\beta }(x,t)=2\arg \left[ \frac{-\sinh \varphi +i\cos \theta 
}{\cosh \varphi +\sin \theta }\right] -i\ln \left[ \frac{\sinh ^{2}\varphi
+\cos ^{2}\theta }{\left( \cosh \varphi +\sin \theta \right) ^{2}}\right] ,
\label{solSG}
\end{equation}%
where we abbreviated $\varphi =\beta x+t/\beta $. Using the relation between
the argument function and the $\arctan $ function equation (\ref{solSG}) can
be converted into the more practical form%
\begin{equation}
\phi _{i\theta ;\beta }(x,t)=\left\{ 
\begin{array}{ll}
4\arctan \left[ \frac{\sqrt{\sinh ^{2}\varphi +\cos ^{2}\theta }+\sinh
\varphi }{\cos \theta }\right] -i\ln \left[ \frac{\sinh ^{2}\varphi +\cos
^{2}\theta }{\left( \cosh \varphi +\sin \theta \right) ^{2}}\right]  & \quad 
\text{for }\theta \neq \pm \frac{\pi }{2} \\ 
-i\ln \left[ \frac{\sinh ^{2}\varphi }{\left( \cosh \varphi \pm 1\right) ^{2}%
}\right]  & \quad \text{for }\theta =\pm \frac{\pi }{2}%
\end{array}%
\right. .  \label{SGsol}
\end{equation}%
The real part is $\mathcal{PT}_{-}^{(\pm )}$-symmetric and the imaginary
part respects a $\mathcal{PT}_{-}^{(0)}$-symmetry, such that overall $\phi
_{i\theta ;\beta }$ is $\mathcal{PT}_{-}^{(\pm )}$-symmetric. As depicted in
figure \ref{SGone} for $\theta \neq \pm \pi /2$ the real part of the
solution constitutes a kink solution accompanied by a one-soliton solution
in the imaginary part. For $\theta =\pm \pi /2$ the real part of the
solution vanishes and the imaginary part becomes a cusp type solution as can
be found for instance in \cite{kawamoto}.

\FIGURE{ \epsfig{file=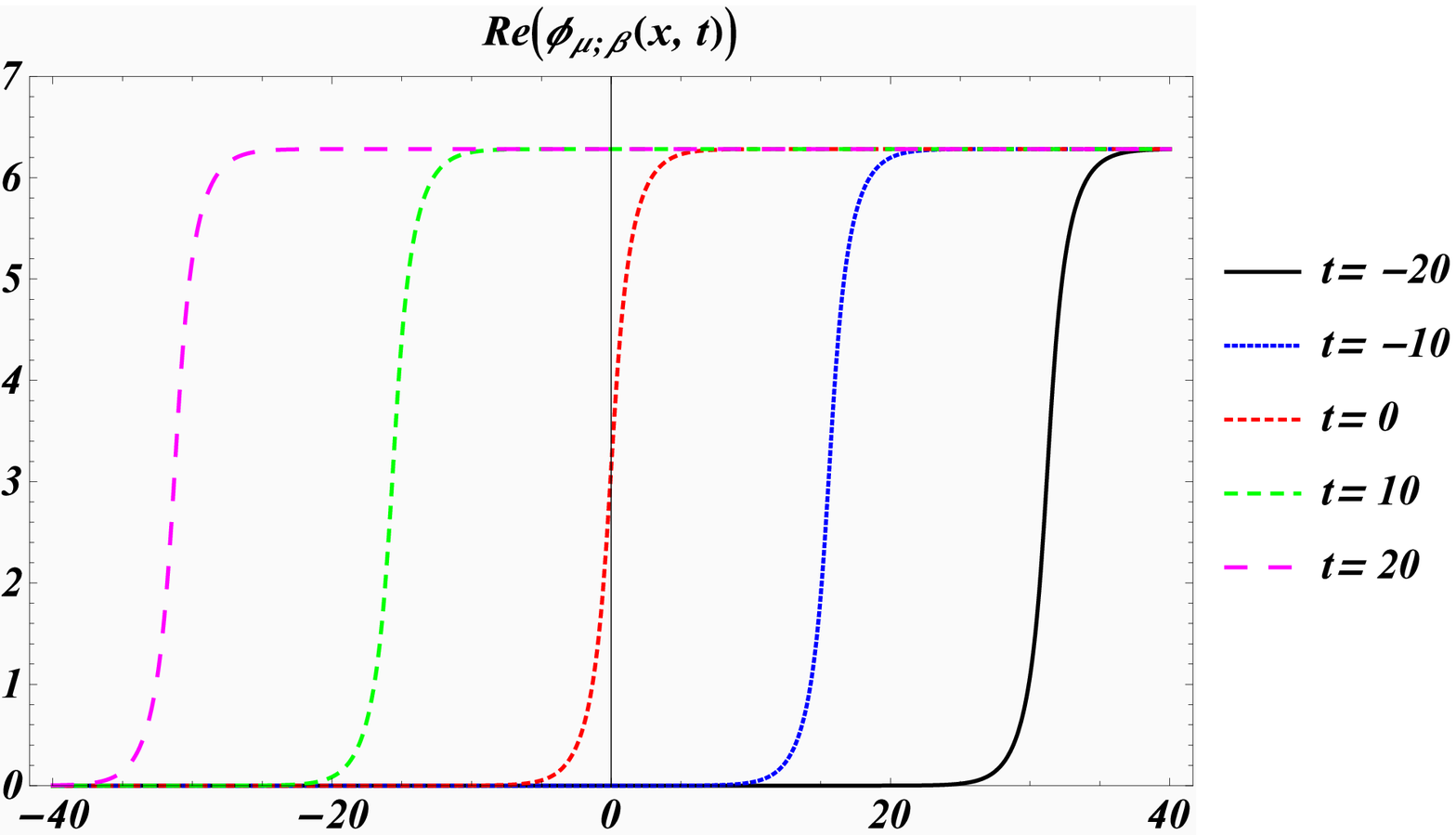,height=4.4cm} \epsfig{file=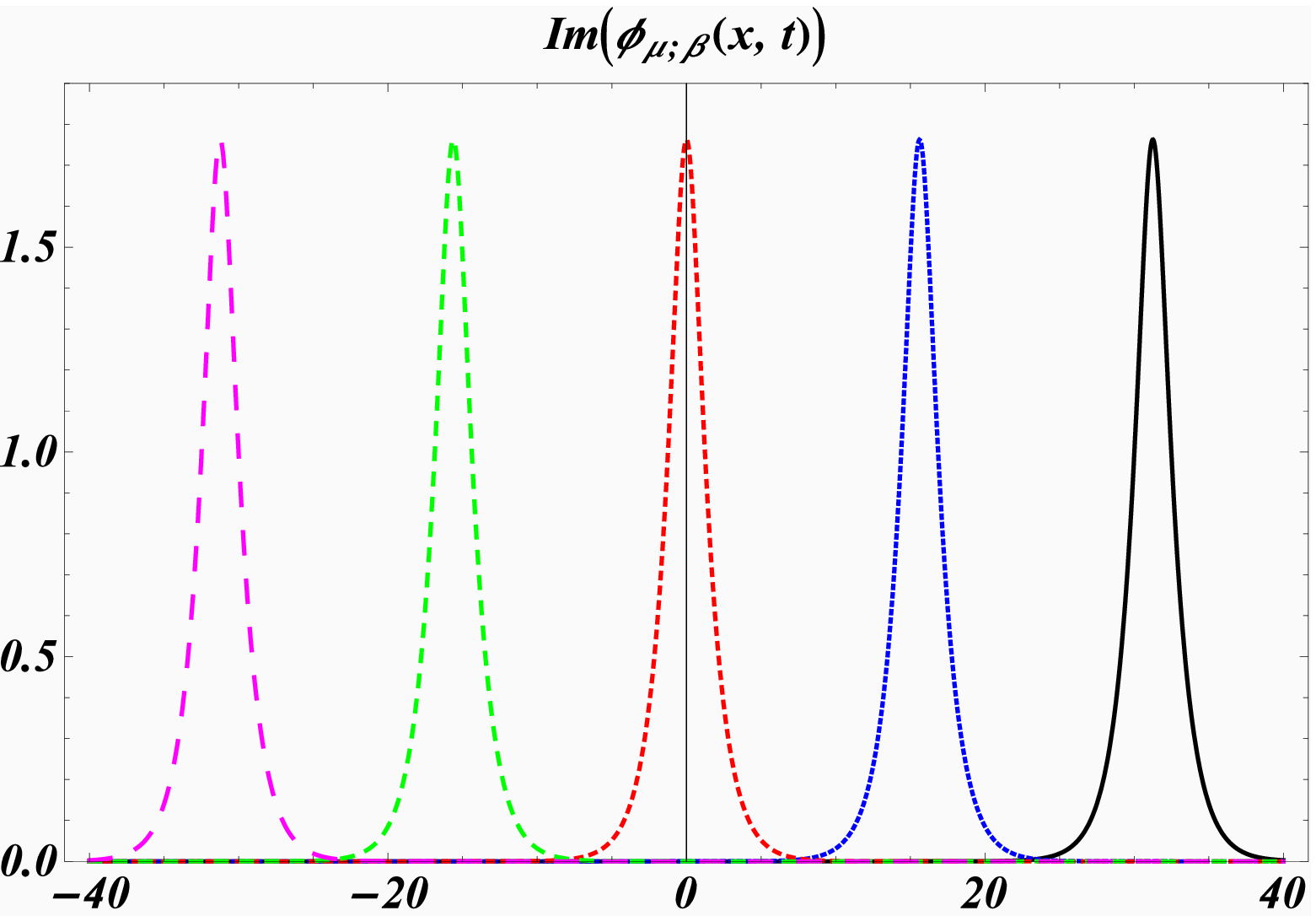,height=4.4cm}
        \caption{Complex sine-Gordon solution with kink as real part and soliton as imaginary part at different values of time for $\alpha=4/5$ and $\mu=i \pi/4$.}
        \label{SGone}}

Let us now construct a two-soliton solution from the sine-Gordon complex
solitons using the B\"{a}cklund transformation which associates two
different types of solutions $\phi $ and $\phi ^{\prime }$ via the two
solutions%
\begin{equation}
\frac{\phi _{x}+\phi _{x}^{\prime }}{2}=\frac{1}{\kappa }\sin \left[ \frac{%
\phi _{x}-\phi _{x}^{\prime }}{2}\right] ,\qquad \text{and\qquad }\frac{\phi
_{t}-\phi _{t}^{\prime }}{2}=\kappa \sin \left[ \frac{\phi _{x}+\phi
_{x}^{\prime }}{2}\right] .  \label{BT}
\end{equation}%
In this case the \textquotedblleft nonlinear superposition
principle\textquotedblright\ relates four solutions $\phi _{0}$, $\phi _{1}$%
, $\phi _{2}$, $\phi _{3}$ as 
\begin{equation}
\tan \left[ \frac{\phi _{3}-\phi _{0}}{4}\right] =\frac{\kappa _{1}+\kappa
_{2}}{\kappa _{1}-\kappa _{2}}\tan \left[ \frac{\phi _{1}-\phi _{2}}{4}%
\right] .  \label{super}
\end{equation}%
Taking now $\phi ^{\prime }=0$, $\phi =\phi _{\mu ;a}$ we identify the
constant in (\ref{BT}) as $\kappa =1/\alpha $. Then taking $\phi _{1}=\phi
_{\mu ;a}$, $\phi _{2}=\phi _{\nu ;\beta }$ and $\phi _{3}=\phi _{\mu ,\nu
;\alpha ,\beta }$ equation (\ref{super}) leads to the new complex
two-solution solution%
\begin{equation}
\phi _{\mu ,\nu ;\alpha ,\beta }=4\arctan \left[ \frac{\beta +\alpha }{\beta
-\alpha }\tan \left( \frac{\phi _{\mu ;a}-\phi _{\nu ;\beta }}{4}\right) %
\right] .
\end{equation}%
This solution exhibits the same kind of symmetry properties as the
one-soliton solutions as we observe in figure \ref{SGtwo}. When $\mu \neq
i\pi /2$, $\nu \neq i\pi /2$ the real part consists of a kink-kink
scattering and the imaginary part of a two soliton scattering.

\FIGURE{ \epsfig{file=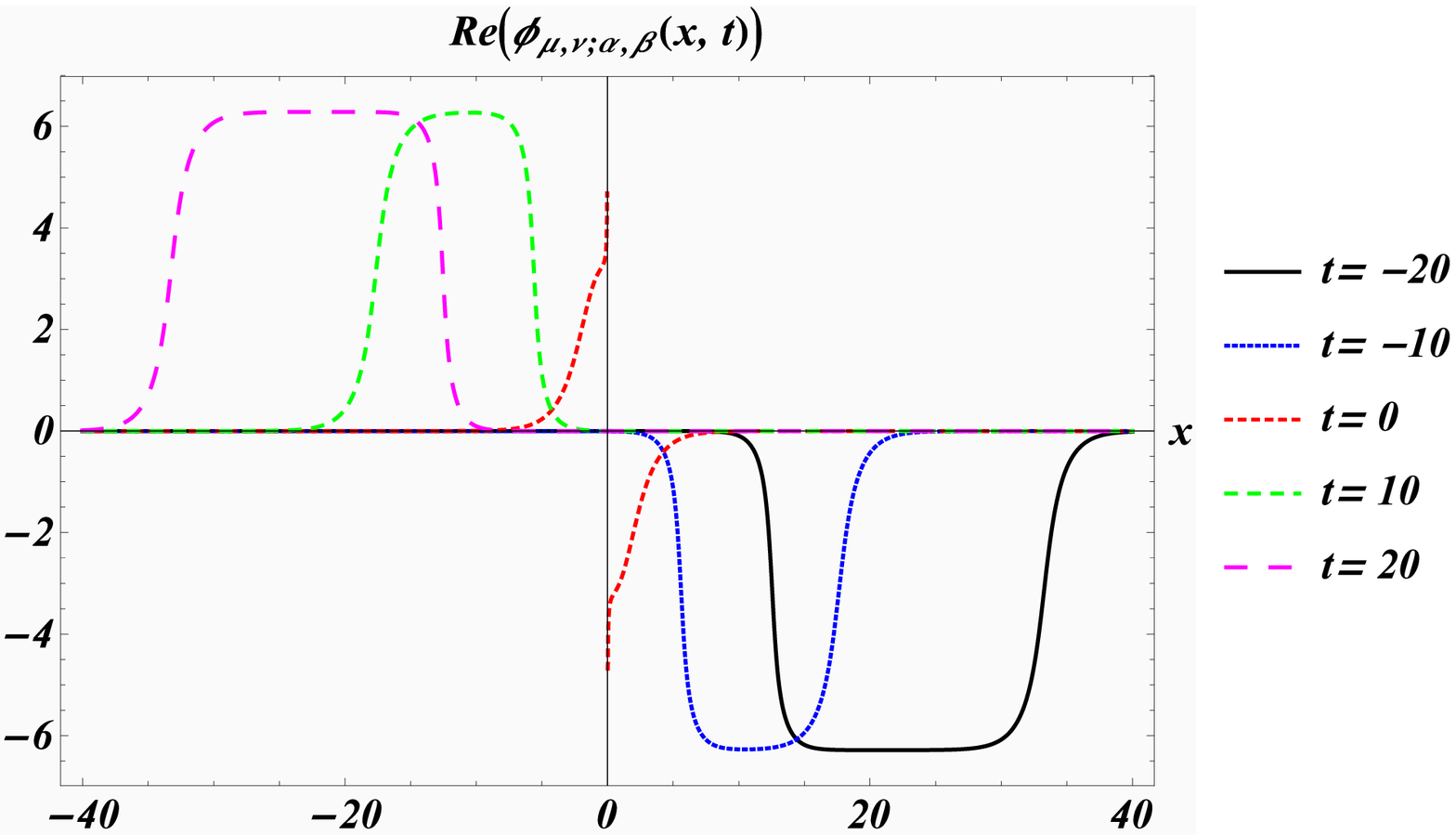,height=4.5cm} \epsfig{file=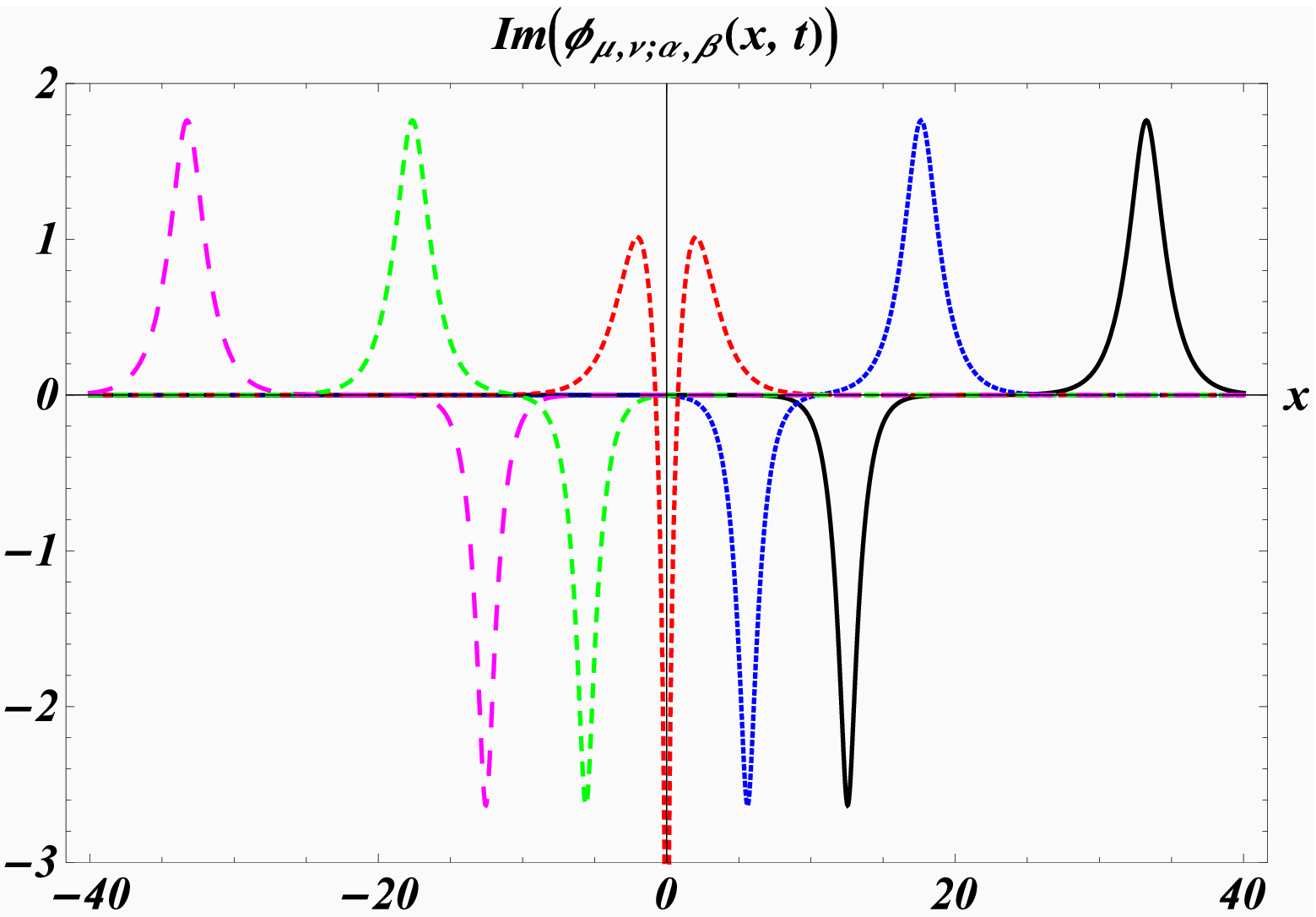,height=4.5cm}
        \caption{Complex sine-Gordon solution with kink-kink scattering in the real part and soliton-soliton scattering in the imaginary part at different values of time for $\alpha=6/5$, $\beta=4/5$, 
$\mu=i \pi/3$   and $\nu=i \pi/4$.}
        \label{SGtwo}}

As computed in (\ref{SGsol}), when $\mu =i\pi /2$ the kink solution in the
real part vanishes and the soliton solution in the imaginary part
degenerates into a cusp. Choosing $\mu =\nu =i\pi /2$ we observe a two cusps
scattering in the imaginary part. These features are depicted in figure \ref%
{SGcusp}.

\FIGURE{ \epsfig{file=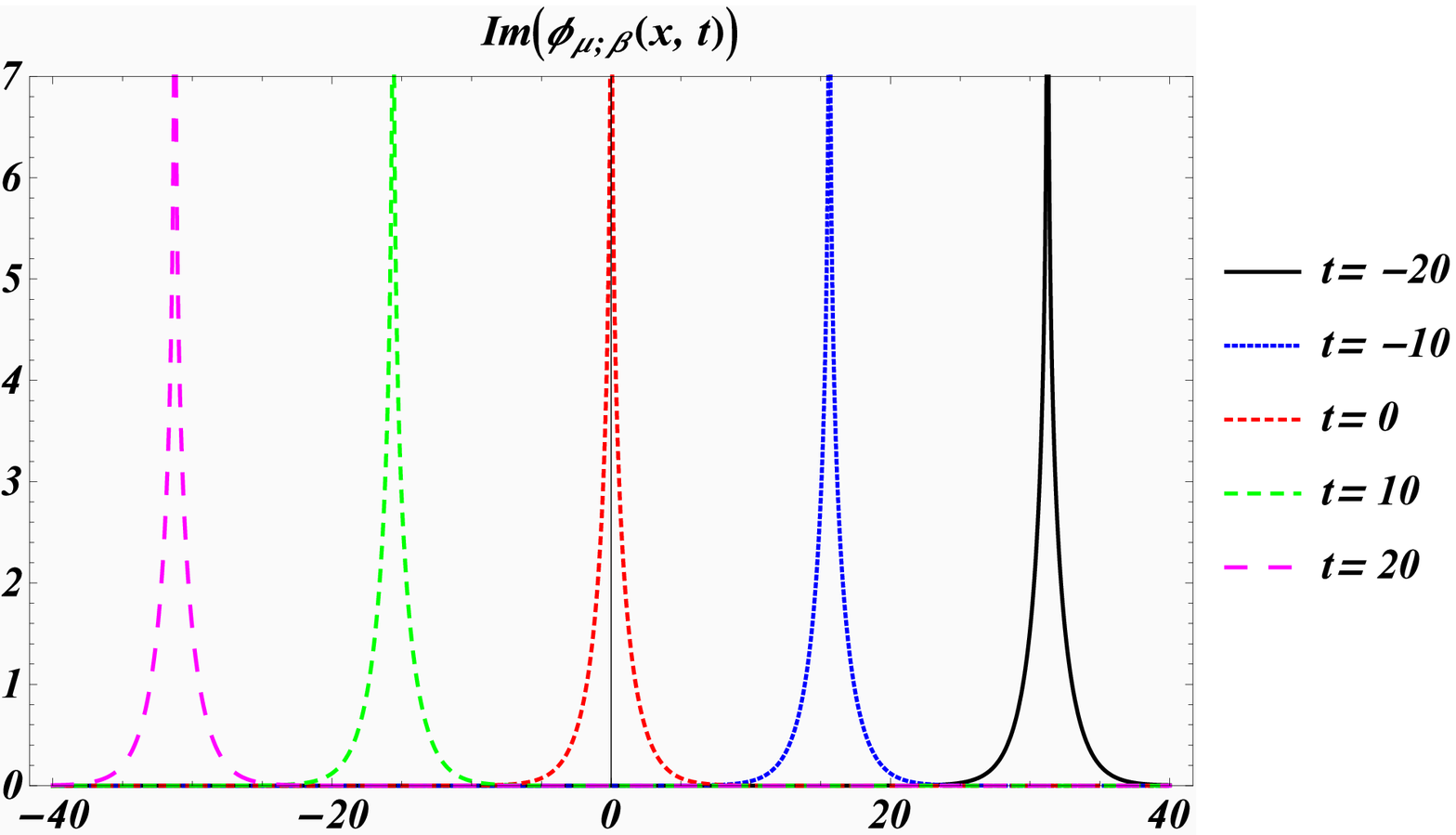,height=4.5cm} \epsfig{file=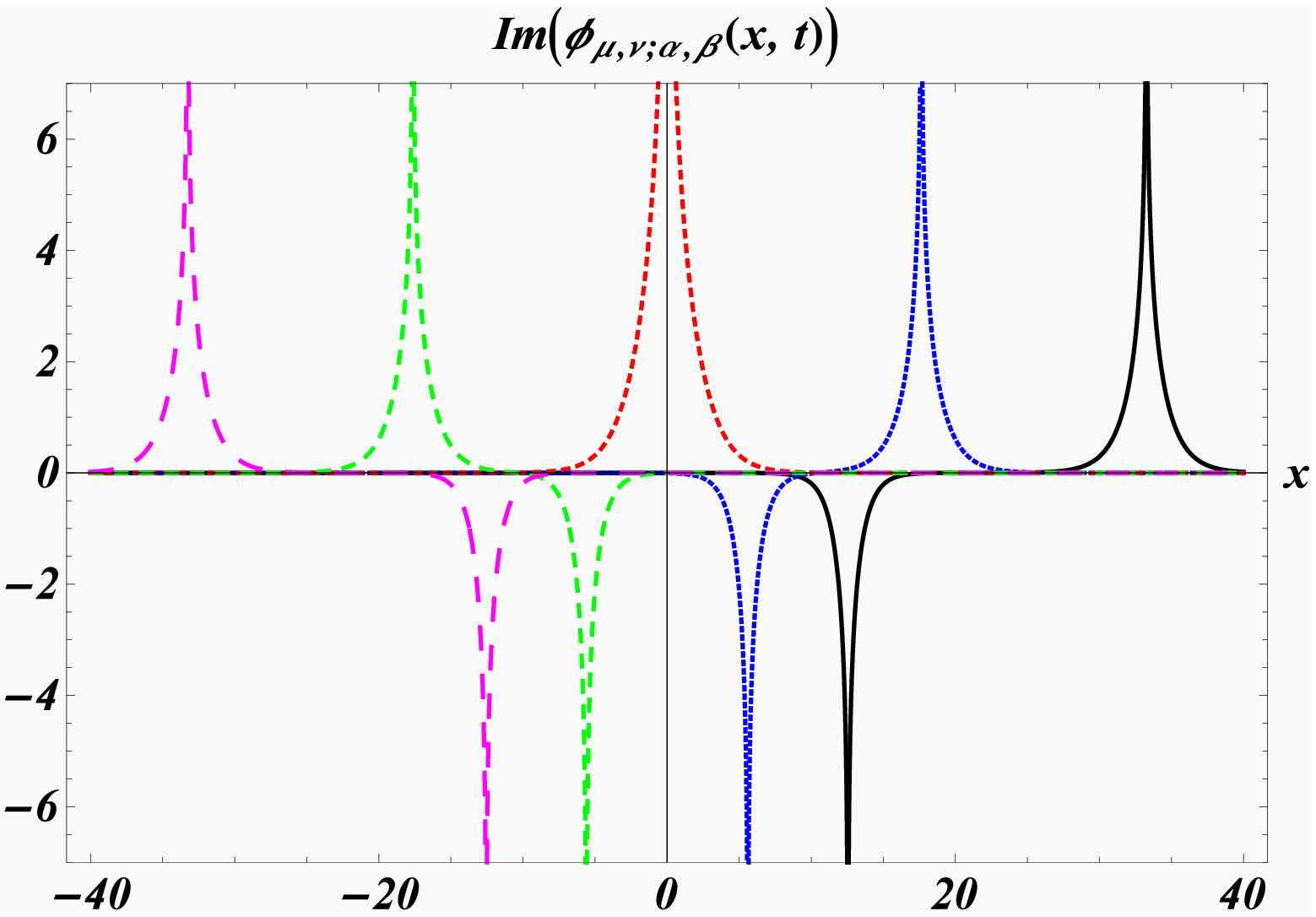,height=4.5cm}
        \caption{Complex sine-Gordon travelling one-cusp solution with $\beta=4/5$, $\mu=i \pi/2$  and two-cusp scattering solution for $\alpha=6/5$, $\beta=4/5$ and 
$\mu=i =\nu =i \pi/3$  at different values of time.}
        \label{SGcusp}}

\subsubsection{Real energies from $\mathcal{PT}$-symmetric and broken $%
\mathcal{PT}$-symmetric solutions}

The Hamiltonian density for the sine-Gordon equation is well-known, see e.g. 
\cite{rajar}. When converted to light-cone variables it reads 
\begin{equation}
\mathcal{H}(\phi ,\phi _{x},\phi _{t})=\frac{1}{4}\left( \phi _{x}^{2}+\phi
_{t}^{2}\right) +1-\cos (\phi ).
\end{equation}%
From this expression we compute real energies for all times $t$ and any
values $\mu $ to%
\begin{equation}
E_{\mu ;\beta }^{SG}=\int\limits_{-\infty }^{\infty }\mathcal{H}\left[ \phi
_{\mu ;\beta },(\phi _{\mu ;\beta })_{x},(\phi _{\mu ;\beta })_{t}\right] dx=%
\frac{(1+\beta ^{2})^{2}}{\beta ^{2}}\int\limits_{-\infty }^{\infty }\func{%
sech}^{2}(\beta x+t/\beta +\mu )dx=\frac{2(1+\beta ^{2})^{2}}{\beta ^{3}}.
\end{equation}%
Once again the imaginary parts of $\mathcal{H}$ do not contribute as the are
already or, by suitable shifts, can be made $\mathcal{PT}$-symmetric.
Numerically we also confirm that the energy of the two-soliton solution is
the sum of the individual one-soliton solutions%
\begin{equation}
E_{\mu ,\nu ;\alpha ,\beta }^{SG}=\int\limits_{-\infty }^{\infty }\mathcal{H}%
\left[ \phi _{\mu ,\nu ;\alpha ,\beta },(\phi _{\mu ,\nu ;\alpha ,\beta
})_{x},(\phi _{\mu ,\nu ;\alpha ,\beta })_{t}\right] dx=E_{\mu ;\alpha
}^{SG}+E_{\mu ;\beta }^{SG},
\end{equation}%
at all times $t$ and any values of $\mu $ and $\nu $.

\section{Conclusions}

Using various techniques, such as Hirota's direct method, B\"{a}cklund and
Miura transformations, we have constructed complex one and two-soliton
solutions to the complex KdV, mKdV and sine-Gordon equations. Some of the
solutions turned out to be $\mathcal{PT}$-symmetric, whereas others have
broken $\mathcal{PT}$-symmetry, as for instance the two-soliton solution
obtained from Hirota's method. Nonetheless, despite the fact that the
corresponding Hamiltonian densities are non-Hermitian, all solutions were
found to lead to real energies. While this was to be expected \cite{AFKdV}
for the $\mathcal{PT}$-symmetric solution, it is less obvious why this
should be the case for the broken scenario. However, as we have shown any of
our one-soliton solution may be converted into a $\mathcal{PT}$-symmetric
one-soliton solution by suitable shifts in time or space and any of our
two-soliton solution may be converted into a $\mathcal{PT}$-symmetric
two-soliton solution by suitable simultaneous shifts in time and space.
Since the value of the energy is insensitive to any of these shifts it must
therefore be real. Moreover, when considering the asymptotic behaviour of $N$%
-soliton solutions we conjecture that one might be able to use of the fact
that they separate into $N$ different one-soliton solutions with possible
shifts in time, with each of them contributing a real value to the overall
energy. As we have seen in section 2.1.2 this is certainly correct for the
KdV two-soliton solution when $\mu =\nu =i\pi /2$, but in order to establish
this in more generality we need to investigate in more detail the effect of
the time-delay the for different values of $\mu $ and $\nu $\ and especially
the cases $N>2$ \cite{timedelay}.

The above mechanism explains well why certain complex soliton solutions
posses real energies. Here we have not allowed complex dispersion relations,
i.e. keeping our parameters $\alpha $, $\beta $ real, or permitted complex
parameters occurring directly in the nonlinear wave equations. In fact, also
for those scenarios it was found \cite{CFB} that broken $\mathcal{PT}$%
-symmetric solutions with non-Hermitian Hamiltonian densities may lead to
real energies, although in a much more constrained setting. The mechanism
responsible for the reality of the energy in those cases is still unclear,
but we believe that the studies presented here will also shed light onto
those situations.

\newif\ifabfull\abfulltrue

\end{document}